\documentclass{article} %arxiv
\usepackage{graphicx}
\usepackage{todonotes}

\usepackage{amsmath}
\usepackage{amssymb}
\usepackage{amsfonts}
\usepackage{csquotes}
\usepackage{mathtools}
\usepackage{algorithm}
\usepackage{algpseudocode}
\usepackage[mode=image|tex]{standalone}
\usepackage{tikz}
\usepackage{tikz-3dplot}
\usetikzlibrary{3d,perspective,calc,matrix,positioning}
\usepackage{pgfplots}
\usepackage{pgfplotstable}
\usepgfplotslibrary{units} % Allows to enter the units nicely
\usetikzlibrary{pgfplots.statistics}
\usepackage{xintexpr}
\usepackage{hyperref}
\usepackage{forest}
\usepackage[capitalize,nameinlink]{cleveref}
\usepackage[mode=text,group-four-digits=true,group-separator={,},exponent-product=\cdot,round-mode=places,round-pad=false]{siunitx}
\usepackage{animate}

\DeclarePairedDelimiter{\norm}{\lVert}{\rVert}
\DeclarePairedDelimiter{\abs}{\lvert}{\rvert}
\DeclareMathOperator*{\argmin}{arg\,min}

\providecommand{\ra}{\ensuremath{\Rightarrow}}

\providecommand{\lambdify}[1]{\ensuremath{\hat{\$}(#1)}}

\providecommand{\complexity}[1]{\ensuremath{\mathcal{O}(#1)}}

\providecommand{\thingitenk}{\texttt{Thingi10K}}
\providecommand{\blosc}{\texttt{blosc2}}

\newcommand{\cP}{\ensuremath{\mathcal{P}}}
\newcommand{\cT}{\ensuremath{T}}

\pgfplotsset{compat=1.18,table/search path={.,./data/,../data/},}
\tikzset{
  octree_t/.style={
    orange,
  },
  omnione_t/.style={
    teal,
  },
  plain_t/.style={
    cyan!60!black,
  },
  downsplit_t/.style={
    lime!80!black,
  },
  openvdb_t/.style={
    red,
  },
  normlabel/.style={
    font=\small,
    anchor=west,
    xshift=0.5ex,
    inner sep=0.7ex,
  },
}
\pgfplotsset{compat=1.18}
\pgfplotsset{
  cloudplot/.style={
    xmin=6e-7,
    xmax=1.5,
    grid=major,
    yminorgrids,
    xlabel={Wavelet compression threshold $\varepsilon$},
    table/search path={./cloud/,./data/cloud/,../data/cloud/},
    table/x=threshold,
    table/col sep=comma,
    table/comment chars=\#,
  },
  discard if not/.style 2 args={% cf. https://tex.stackexchange.com/a/572616
    filter discard warning=false,
    x filter/.append code={
      \edef\tempa{\thisrow{#1}}%
      \edef\tempb{#2}%
      \ifnum\pdfstrcmp{\tempa}{\tempb}=0
      \else
          \def\pgfmathresult{inf}%
      \fi
    }
  },
  octree/.style={
    octree_t,every mark/.append style={fill=orange!80!black,solid},mark size=1.5pt,mark=square*
  },
  omnione/.style={
    omnione_t,every mark/.append style={fill=teal!80!black,solid},mark=diamond*
  },
  plain/.style={
    plain_t,every mark/.append style={fill=cyan!80!black,solid},mark=otimes*
  },
  downsplit/.style={
    downsplit_t,every mark/.append style={fill=lime!80!black,solid},mark=diamond*
  },
  openvdb/.style={
    openvdb_t,densely dashed,every mark/.append style={solid,fill=red!80!black},mark=star
  },
  rawvariant/.style={solid},
  bloscvariant/.style={densely dotted},
  can filtered/.style={
    discard if not={variant}{can},
    plain,
  },
  ds filtered/.style={
    discard if not={variant}{ds},
    downsplit,
  },
}

\usepackage{ifthen}
\newboolean{arxiv}
\setboolean{arxiv}{true} %arxiv?
\ifthenelse{\boolean{arxiv}}{
    \usepackage{amsthm}
    \usepackage[style=numeric,maxbibnames=4,maxcitenames=2,backend=biber]{biblatex}
    \addbibresource{bib.bib}
    \AtEveryBibitem{%
    \clearfield{urlyear}%
    }
    \newtheorem{definition}{Definition}[section]

}{
}
\usepackage{subcaption}
\DeclareSIUnit[number-unit-product = ]{\xtimes}{×}%×

\ifthenelse{\boolean{arxiv}}{
\title{Towards Fully Dynamic Omnitrees: Moment-Conserving Anisotropic Compression With Wavelets}
\author{Theresa Pollinger, Masado Ishii, Jens Domke}
\date{RIKEN Center for Computational Science, Kobe, Japan}
}{
\title{Towards Fully Dynamic Omnitrees: Moment-Conserving Anisotropic Compression With Wavelets\thanks{Submitted to the editors
July 2026.}}
\author{Theresa Pollinger\thanks{RIKEN Center for Computational Science, Kobe, Japan (\email{theresa.pollinger@a.riken.jp}, \email{masado.ishii@riken.jp}, \email{jens.domke@riken.jp})\thanks{Chair for Scientific Computing, Universität Stuttgart}}
\and Masado Ishii\footnotemark[2]
\and Jens Domke\footnotemark[2]}
\headers{Moment-Conserving Anisotropic Compression With Wavelets}{Theresa Pollinger, Masado Ishii, and Jens Domke}
}

\begin{document}

\maketitle

\begin{abstract}%reduced to 250 words now!
  Recently, omnitrees were introduced as a flexible space partitioning tree that improves upon the benefits of both octrees and k-d trees:
  Omnitrees' efficient encoding of anisotropic refinements holds particular interest for applications with anisotropic features and high dimensionality.
  These include, but are not limited to, computer graphics, databases, machine learning, and physics simulations.
  The present paper defines new operations on the omnitree encoding that extend its capabilities from the existing refinement to also include \emph{coarsening} and therefore \textit{fully adaptive compression}.
  It demonstrates natural integration of omnitrees with wavelets, which conserves moments of the stored function by design.
  For omnitrees, the wavelet coefficients can be interpreted as local refinement priorities, which can be used to guide the adaptation process.
  We derive algorithms for coarsening and downsplit that are guided by wavelet coefficients, and show their application to a large dataset of 3D shapes, as well as the continuous-valued density field of a cloud.
  The comparison to OpenVDB, a widely-used data structure for sparse volumetric data in computer graphics, enables a demonstration of the practical benefits of omnitrees even for moderately anisotropic three-dimensional data.
  Compared to OpenVDB, objects can be stored using up to \qty[round-precision=0,number-unit-product = ]{27.6145849495733}{\xtimes} less space, and asymptotically show savings that exceed theoretical expectations.
  Using lossy compression, the cloud dataset can be compressed by $\approx5\times$ compared to OpenVDB, with negligible loss of visual quality.
  This demonstrates the potential of omnitrees for efficient storage and processing, and motivates further research into their applications in various domains.
\end{abstract}

\ifthenelse{\boolean{arxiv}}{
}{
%  	65D15: Algorithms for approximation of functions, 68P05: data structures, 65M50: Mesh generation, refinement, and adaptive methods for boundary value problems involving PDEs, 05C05: trees, 65T60: Numerical methods for wavelets, 68P30: Coding and information theory (compaction, compression, models of communication, encoding schemes, etc.), 65D18: Numerical aspects of computer graphics, image analysis, and computational geometry
}

\section{Introduction}\label{sec:introduction}

% Outline
% - previous work: sensitivity analysis for dimension-aware refinement
% - also showed: probabilistic approach not great for practical applications (-> missing wing)
% - practical applications usually: start with finely / "fully" resolved function, then coarsen / compress (e.g. OpenVDB)
% - but how for omnitrees? -> coarsening algorithm, wavelets provide dimension-wise feature coefficients
% - coarsening an omnitree: needs coarsening stacks
% - downsplit=de-normalization in omnitree to allow for more coarsening
% - wavelet trees (maybe reference some Besov / optimality things?)
% - Haar wavelets with binary coefficients perfectly fit the example of binary shape representation on omnitrees
% - heuristic based on downsplit
% - example: rotated blocks
% - here, use the same methodology as before (4166 thingies) to compare to OpenVDB
% - compare: storage used for topology information, storage used for function coefficients
% - future work: forest-of-omnitrees-of-omnitrees-of-omnitrees...-of-omnitrees
% - generalizes to other compactly supported wavelets (like Alpert multiwavelets)
% future work: omnitree-of-omnitrees-...of-omnitrees

Omnitrees are introduced by \cite{pollingerBeautyAnisotropicMesh2025} as a type of flexible space partitioning tree that improves upon the benefits of both octrees and bintrees (k-d trees).
This is achieved by efficiently encoding anisotropic refinements.
Whereas octrees bisect all dimensions of a volume at once, and bintrees bisect one dimension at a time, omnitrees bisect a locally optimal subset of dimensions on any level.
It is shown in \cite{pollingerBeautyAnisotropicMesh2025} that, for the same error threshold, omnitree discretizations of anisotropic problems are smaller than their octree counterparts by a power law, $N^{\textrm{omni}} \sim \left( N^{\textrm{oct}} \right)^\gamma,\,\gamma < 1$, where $N$ is the number of cells in the discretization.
That is, $N$ is improved by faster than a constant factor.
Omnitrees also generalize bintrees to allow any dimension to be split on any level without a regular period.
The ability for multiple dimensions to be split at once further distills the tree into a shallower encoding with shorter traversal depth.

The present paper defines new operations on this omnitree encoding that enable not only top-down \textit{refinement} but also \textit{fully adaptive compression}.
The compression scheme integrates naturally with wavelet bases:
Wavelet coefficients below a threshold indicate that coarsening is possible, and this paper describes how the updated tree and wavelet coefficients can be computed.
For the purposes of this work, we select Haar wavelets as one of the simplest possible multiscale bases.
Haar wavelet compression conserves the mass of the function, and higher orders of conservation (momentum, energy, ...) are possible with a choice of higher-order wavelets.
By comparing with OpenVDB---a state-of-the-art adaptive hierarchical storage format---this paper provides a validation of increased compression and even approximation rate for three-dimensional objects.

These coarsening operations represent a major step towards a fully dynamic omnitree data structure that can be used for adaptive storage and processing of anisotropic data in various domains, including computer graphics, databases, machine learning, and physics simulations.

% say more here...
% - compression = dropping wavelet coaefficients and coarsening the tree accordingly
% -> good conservation and other properties, depending on choice of wavelets
% - here, simplest choice: Haar wavelets 
% - validate with discrete (3d objects)- and continuous valued functions (cloud)
% - compare to OpenVDB

\section{Related Work}\label{sec:related_work}
In addition to the related work of the 2000s, e.g.~\cite{domelSplitflowProgress3D2000,ogawaAdaptiveCartesianMesh2003}, discussed in the previous omnitree paper~\cite{pollingerBeautyAnisotropicMesh2025}, there is a range of related concepts that the authors since became aware of.
On the topic of omnitree-like data structures, there is a tradition of including anisotropically split hexagonal elements in mixed element trees~\cite{hitschfeldMixedElementTrees1993,holkeT8codeModularAdaptive2025}.
Also using mixed-element trees, the $hp\mathrm{3D}$ library~\cite{hennekinghp3DScalableMPI2024} was used successfully to explore the numerics of Discontinuous Petrov-Galerkin schemes on omnitree-equivalent dyadic discretization spaces~\cite{chakrabortyAnisotropicHpadaptationFramework2024a}.
Their $hp$ adaptation approach, in particular the anisotropic $p$ adaptation, provides a very interesting possibility to further increase the convergence order of omnitree discretizations for physics simulations.
Similarly, the interfaces for MFEM\footnote{\url{https://web.archive.org/web/20240811233204/https://mfem.org/howto/ncmesh/}} support anisotropic refinement of hexahedral elements~\cite{cervenyNonconformingMeshRefinement2019}.
The same holds true for the interface of peano in the refinement\footnote{\url{https://web.archive.org/web/20250918090327/https://hpcsoftware.pages.gitlab.lrz.de/Peano/dd/dc6/structpeano4_1_1datamanagement_1_1VertexMarker.html\#a5bd3c7eec8d87ec4d2d403639063c21c}} of its triadic spacetree data structure~\cite{weinzierlPeanoTraversalStorage2011}.

% maybe also Alauzet / metric-based anisotropic meshing~\cite{loseille}
% maybe difference between LBVH and omnitree?
OpenVDB~\cite{musethVDBHighresolutionSparse2013} is a hierarchical voxel structure specialized for $3$-d animation tasks such as volume rendering and finite difference calculations.
Although the authors acknowledge a \enquote{superficial} resemblance to octree grids, they emphasize properties of the data structure, such as compile-time-fixed tree height and traversal caching, that enable fast lookups despite its dynamic nature.

Since the present work draws some comparisons to OpenVDB, we would also like to mention recent approaches to increase its parallelism and information efficiency.
For example, NanoVDB~\cite{musethNanoVDBGPUFriendlyPortable2021} was created to offload a static snapshot of an OpenVDB tree (similar to the linearized omnitree data structure) to various GPU architectures.
This is particularly useful for real-time applications~\cite{walkerNanoMapGPUAcceleratedOpenVDBBased2022} and allows for further lossy compression to achieve fast scientific visualization~\cite{zellmannGPUVolumeRendering2025}.
The hierarchical lossy compression~\cite{zellmannGPUVolumeRendering2025} has some similarity to the lossy compression that this paper presents in \cref{sec:compression:coarsening}.

NeuralVDB~\cite{kimNeuralVDBHighresolutionSparse2024a} uses learned details to achieve a \qtyrange{10}{100}{\xtimes} compression compared to OpenVDB on various surfaces.
(In fact, one of their showcases is the same cloud that will be the experimental basis for \cref{sec:results:cloud}.)
One can argue that these findings are not in conflict with omnitrees, to the contrary:
Omnitree compression may well benefit from neural approaches, potentially compounding the compression effects of both approraches.
Recently, NeuralVDB approaches were integrated into $f$VDB, which---among other optimizations---allows storing the function data separately from the tree topology data~\cite{williamsFVDBDeepLearningFramework2024}.
% - could also be suitable for the space-frequency tiling algorithms explored by Villemoes et al.~\cite{villemoes} -> in discussion

\section{Omnitree Transformations Beyond Refinement}\label{sec:omnitree_operations}
Before introducing the novel transformations coarsening (\cref{sec:compression:coarsening}) and downsplit (\cref{sec:compression:downsplit}), we briefly recapitulate the core concepts of omnitrees~\cite{pollingerBeautyAnisotropicMesh2025}:

Omnitrees partition a $d$-dimensional Cartesian domain $\Omega \subset \mathbb{R}^d$ into $N$ non-overlapping subdomains,
\begin{equation}
\overline{\Omega} = \bigcup_{1 \leq i \leq N} \overline{Q_i}, \quad Q_i \cap Q_j = \emptyset \quad \text{for } i \ne j,
\end{equation}
where $Q = (x_1^-, x_1^+) \times \cdots \times (x_d^-, x_d^+)$ are $d$-dimensional hyper-rectangles (hereafter \enquote{rectangles}, or \enquote{cuboids} in 3-d).
Indexing dimensions by the set $\mathcal{D}$ with $\lvert\mathcal{D}\rvert = d$, we take $\Omega = (0,1)^\mathcal{D}$ without loss of generality.
An omnitree encodes a hierarchical dyadic partition of this unit hypercube: all partition boundaries occur at dyadic rationals, i.e., at multiples of $2^{-k}$ for appropriate $k \in \mathbb{N}$.

An omnitree node $v$ is characterized by its binary splitting label $\vec{b}(v) \in \{0,1\}^d$, where $b_j = 1$ indicates bisection in dimension $j$, yielding $2^{|\sigma(v)|}$ children.
The all-zero label $\vec{b} = \vec{0}$ is reserved for leaf nodes.
Normalization~\cite[Definition 2.2]{pollingerBeautyAnisotropicMesh2025} enforces that splits are gathered as high in the tree as possible, guaranteeing a unique, shallowest representation.

Each node maps to a rectangle $Q_{\vec{i},\vec{\ell}}$ with per-dimension refinement levels $\vec{\ell} \in \mathbb{N}^d$, extents $2^{-\vec{\ell}}$, and index $\vec{i}$ with $i_j \in \{0,\ldots,2^{\ell_j}-1\}$, occupying
\begin{equation}
Q_{\vec{i},\vec{\ell}} = \prod_{j=1}^d \bigl(i_j 2^{-\ell_j},\, (i_j+1)2^{-\ell_j}\bigr).
\end{equation}
% Rectangles can be identified by AMM location codes $\stringify{\cdot}$~\cite{bhatiaAMMAdaptiveMultilinear2022}, extended to $\lambdify{\cdot}$ by appending $\lambda$ in each split dimension.
The tree is linearized into a \emph{binary descriptor} via Z-curve order, allowing to access a node $v^i$'s splitting label $b(v^i)$ by accessing the descriptor's $i$th tuple of $d$ bits.
If spatial data is stored on leaves, it can be stored in a flat \emph{data array} the same length as the number of leaves.

Refinement proceeds in four steps:
\begin{enumerate}
    \item \emph{Marker attachment}: Each node $v$ in the initial tree receives a marker $\vec{m}(v) \in \mathbb{N}^d$, where $m_j$ specifies additional refinement levels in dimension $j$; the default is $\vec{m}(v) = \vec{0}$
    \item \emph{Bottom-up sweep}: Markers migrate upward when all siblings share a common marker or already carry a split ($b_j=1$) in the relevant dimension; the latter are compensated with negative markers.
    \item \emph{Top-down sweep}: Markers lifted too high are pushed back down to children where the refinement can be resolved. 
    \item \emph{Tree construction}: 
    The target tree is built with preorder depth-first traversal, greedily constructing the target descriptor.
    At each rectangle $q$, the initial node $s$ whose rectangle covers $q$ is located.
    Child rectangles $w_e$ are identified by substituting trailing $\lambda$ entries in $\lambdify{v}$ with $0$ or $1$, and the procedure recurses.
    % \todo{Stringification/lambdification not further mentioned -> commented}
    At leaf nodes, the subtree is expanded directly from $\vec{m}$.
    Structural changes can be substantial: positive markers increase branching (requiring grandchild adoption) and negative markers decrease it (requiring node culling).
\end{enumerate}

The algorithm guarantees a valid omnitree, though not necessarily in its unique normalized representation;
The latter can be obtained in a separate step by detecting nodes that violate the uniqueness condition and attaching the corresponding markers to the node (positive marker) and its children (negative markers) and applying the tree construction step.
This is repeated until no nodes in the newly constructed tree violate the uniqueness condition.

\subsection{Information Mapping}\label{subsec:mapping}
An index \emph{mapping} $M$ from an initial tree $\cT$ (length~$N$) to a
target tree $\cT'$ (length~$N'$)
\begin{equation}\label{eq:M}
  M : \{0,\dots,N{-}1\} \;\to\; \cP\!\bigl(\{0,\dots,N'{-}1\}\bigr),
\end{equation}
always maps in the {initial $\to$ target} direction (and \cP{} denotes the power set).
% As a data structure, it can be used to track where function values need to move when going from the initial to the target discretization.
The mapping $M[i]$ for initial node $i$ contains at least a node in the target tree that corresponds to the smallest rectangle in the target discretization that covers $i$'s rectangle in the initial discretization.
Also, there is a union of corresponding target rectangles in $M[i]$ that exactly matches $i$'s initial rectangle.
The mapping is important to the algorithms presented here, as it is an expression of how information is propagated around the tree.

For example, considering refinement as presented in \cite{pollingerBeautyAnisotropicMesh2025}: 
Refining a leaf node~$v^i$ with target refinement $\vec{b}(v^{i'})$ expands it into a parent with $2^{\abs{\vec{b}(v^{i'})}}$ leaf children.
Assuming no further tree modifications happen (for example, all of $v^i$s siblings are leaves too), the mapping is
\begin{equation}\label{eq:refine}
  M[i] = \{i', i'{+}1, \dots, i'{+}2^{\abs{\vec{b}(v^i)}}\},
\end{equation}
where $i'$ is the target index of $v^i$.
All other nodes shift by $\delta = 2^{\abs{\vec{b}(v^{i'})}}$:
$M[j] = \{j + \delta\}$ for $j > i$.
New information is added at the refined leaf $i$, by finer resolving the part of the discretized space it maps to.

For refinement, the mapping can always be directly created, for example as an array of sets, during construction of the target tree as part of the refinement algorithm in \cite{pollingerBeautyAnisotropicMesh2025}.

% decision: not all details, but reasoning behind implementation algorithm logic for mapping / postprocessing of mapping

\subsection{Coarsening}\label{sec:coarsening}

Coarsening an internal node reverses the action of refinement, decrementing selected positions from the splitting label and fusing children along one or more dimensions.
% In its basic form, this operation may leave the tree unnormalized. -> no! only if it was unnormalized before
A request to coarsen the tree consists of a set of \textit{fusion markers} attached to internal nodes.

\begin{definition}[Coarsening]\label{def:coarsening}
Let $v$ be an internal node of a $d$-dimensional omnitree with splitting label $\vec{b}(v)$ and $\sigma(v) = \{j \in \mathcal{D} : b_j(v) = 1\}$.
A coarsening of $v$ in dimensions $C \subseteq \sigma(v)$, $C \neq \emptyset$, replaces $v$ with a node $v'$ whose splitting label satisfies $b_j(v') = 0$ for $j \in C$ and $b_j(v') = b_j(v)$ for $j \notin C$.
The $2^{\abs{\sigma(v)}}$ children of $v$ are fused along each dimension in $C$, yielding $2^{\abs{\sigma(v) \setminus C}}$ children of~$v'$; in the case that $C = \sigma(v)$, $v'$ is a leaf and as such has no children left.
\end{definition}
%what exactly about nodes that become a leaf but are not parents of a leaf? -> don't consider for now

In case $C=\sigma(v)$, all split dimensions at the node are coarsened, and all children are absorbed.
When $C \subsetneq \sigma(v)$, the complement is nonempty ($\bar C \neq \emptyset$) and new fused children persist: 
Specifically, a group of children with common $\bar C$-dimension indices are fused into a single child of~$v'$.

Theoretically, fusion markers placed at arbitrary internal nodes would be conditional on the structure of the tree at lower levels.
For simplicity, we restrict our presentation to last-level coarsening requests, that is, requests to reverse last-level refinements, which are guaranteed to result in the expected fusion.

Beneath a marked node, fusion is well-defined if only if all branches contain identical subtrees over the non-coarsened dimensions.
Equivalently, it must be possible to rotate the coarsened dimensions to the bottom, i.e., to parents of leafs, where coarsening would be a series of local modifications to the tree.
If branches contain identical subtrees, the information mapping is determined by a constant stride of the subtree size.
To compute this stride takes linear time in the size of one of the (identical) subtrees.
This cost is amortized over mapping the entire family of subtrees.

Both conditions---last-level markers and identical subtrees---are met by nodes whose children already consist entirely of leaves.

In practice, coarsening can be implemented as a special case of the refinement algorithm of~\cite{pollingerBeautyAnisotropicMesh2025}.
A coarsening request for node $v$ in dimensions $C$ is expressed as a negative marker $\vec{m}(v)$, with $m_j = -1$ for $j \in C$ and $m_j = 0$ otherwise.
The marker-sweep machinery can operate on these negative markers analogously to positive ones.
In contrast to refinement, where the splitting markers can always be attached to leaves, our code~\cite{bleifreiFreifrauvonbleifreiDyAda2025} attaches coarsening markers directly to internal (parent) nodes where the dimension is currently refined.
Negative markers do not move in the upward and downward sweep phases.
\footnote{Technically, they can be swept down the tree if they cannot be realized at their original node, in the case that $v$ is not initially refined in all of $C$ (though this only applicable when the omnitree is not canonical~\cite[Definition 2.2]{pollingerBeautyAnisotropicMesh2025}, and the refinement in $C$ eventually does occur for all the children nodes).}
% For the purposes of this paper, we always restrict coarsening to the case where all children of the coarsened node are leaves, i.e. their label $b_j$ is zero for all dimensions $j \in 1,\ldots,d$. %already mentioned above

In the construction step, the new splitting label is computed as $\vec{b}(v') = \vec{b}(v) + \vec{m}(v)$, where the addition is component-wise and $b_j \in \{0,1\}$.
If $b_j(v) = 1$ and $m_j(v) = -1$, then $b_j(v') = 0$: that dimension is removed from the split.

From the final splitting labels, a linearized descriptor of the target tree can be constructed.
%When a dimension $j$ is removed from $\sigma(v)$, two children with common indices except the $j$-th index are fused.
%Naively, this can create a problem for the linearized tree construction algorithm, which is designed to construct the target tree greedily, in a single forward pass in Z order.
Consider an initial node $v$ with $2^{\abs{\sigma(v)}}$ children, indexed by tuples $e \in \{0,1\}^{\sigma(v)}$ corresponding to local Morton codes.
Let $C \subseteq \sigma(v)$ be the coarsening request at $v$ and $\bar{C} = \sigma(v) \setminus C$ be the unaffected split dimensions. 
To coarsen $v$ in dimensions $C$, each group of children in the initial tree with common $\bar{C}$ coordinates should be fused into one child of $v'$ in the target tree, the goal being for $v'$ to end with $2^{\abs{\bar{C}}}$ total children.
The tree construction algorithm of~\cite{pollingerBeautyAnisotropicMesh2025} must be modified here.
As stated, it builds the target descriptor top-down by locating, for each child, the corresponding child in the initial tree via its location code stack $\$$.
When it encounters the first member $v_a$ of a yet unfused group in the initial tree, it emits a node $v_a'$ for the target tree.
One can track their connection by adding $a'$ to $M[a]$ in the mapping $M$.
However, further members $v^i$ of the group to be fused would not have their mapping $M[i]$ found---because the algorithm is not looking for them.
In addition to producing an invalid mapping, the naive algorithm cannot know where the subtree region of the initial tree ends.
Even though refinement of the subtree is already finalized and the subtree is greedily added to $T'$, there may be tree nodes that are erroneously copied from the initial tree.
The solution is to find all members of the fused group when their target node $v_a'$ is emitted, by appending their local binary location to $\$$ in line 17 of algorithm 1 in \cite{pollingerBeautyAnisotropicMesh2025}.
A special case has to be considered when all split dimensions are removed, i.e. $\vec{b}(v') = \vec{0}$:
No children of $v'$ are emitted to the target tree in this case.
As a result, all the coarsened children nodes $v^i$, initially leaves, will map to the index of $v'$.
The $v^i$ are easily identified in the construction algorithm, as they are the $2^{\abs{\sigma(v)}}$ direct successors of $v$ in the initial tree's linearized descriptor.

Constructing the mapping this way allows to exactly identify the location of input and output data for the wavelet compression in the initial and target data vectors as will be described in \cref{sec:coarsening_coefficients}.

\subsection{Downsplit}\label{sec:downsplit}

Downsplit restructures an omnitree without modifying the set of leaf boxes.
It factors a multi-dimensional split into a cascade of lower-dimensional splits.
Again, the outputs of the procedure are a target tree and a mapping $M$ from initial to target tree nodes.
\begin{definition}[Downsplit]\label{def:downsplit}
Let $v$ be an internal node of omnitree $\cT$ with splitting label $\vec{b}(v)$ and $\abs{\sigma(v)} \geq 2$.
$v$ has children $u_k$, $k \in \{0,1\}^{\sigma(v)}$.
A downsplit of dimensions $S_1 \subset \sigma(v)$, $\emptyset \neq S_1 \neq \sigma(v)$, with $S_2 = \sigma(v) \setminus S_1$, replaces $v$ by a two-level structure:
\begin{enumerate}
    \item A node $v'$ in the target tree $\cT'$ at the same location, with $\sigma(v') = S_2$.
    Thus $v'$ has $2^{\abs{ S_2}}$ children.
    \item Each (newly added) child $u_r$ of $v'$ ($r \in \{0,1\}^{S_2}$) has $\sigma(u_r) = S_1$.
    Accordingly, $u_r$ has $2^{\abs{ S_1}}$ children, which are exactly the original children $u_k$ whose $S_2$-coordinates equal~$r$.
\end{enumerate} 
\end{definition}

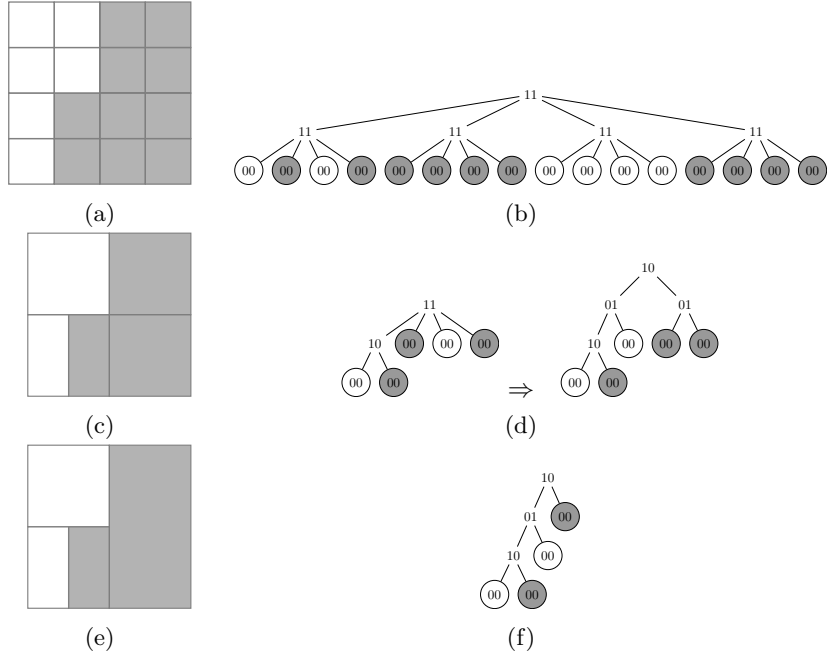
\begin{figure*}
    \centering
    \begin{subfigure}{0.2\textwidth}
        \centering
        \begin{tikzpicture}
\draw[very thin, white] (0,0) rectangle (1,1);
\providecolor{color_0}{RGB}{255,255,255}
\providecolor{color_1}{RGB}{0, 0, 0}
\draw[very thin, gray, fill=color_0,fill opacity=0.3] (0.000000,0.000000) rectangle (0.250000,0.250000);
% \node[inner sep=0] (A) at (0.125,0.125) {};
\draw[very thin, gray, fill=color_1,fill opacity=0.3] (0.250000,0.000000) rectangle (0.500000,0.250000);
% \node[inner sep=0] (B) at (0.375,0.125) {};
\draw[very thin, gray, fill=color_0,fill opacity=0.3] (0.000000,0.250000) rectangle (0.250000,0.500000);
% \node[inner sep=0] (C) at (0.125,0.375) {};
\draw[very thin, gray, fill=color_1,fill opacity=0.3] (0.250000,0.250000) rectangle (0.500000,0.500000);
% \node[inner sep=0] (D) at (0.375,0.375) {};
\draw[very thin, gray, fill=color_1,fill opacity=0.3] (0.500000,0.000000) rectangle (0.750000,0.250000);
% \node[inner sep=0] (E) at (0.625,0.125) {};
\draw[very thin, gray, fill=color_1,fill opacity=0.3] (0.750000,0.000000) rectangle (1.000000,0.250000);
% \node[inner sep=0] (F) at (0.875,0.125) {};
\draw[very thin, gray, fill=color_1,fill opacity=0.3] (0.500000,0.250000) rectangle (0.750000,0.500000);
% \node[inner sep=0] (G) at (0.625,0.375) {};
\draw[very thin, gray, fill=color_1,fill opacity=0.3] (0.750000,0.250000) rectangle (1.000000,0.500000);
% \node[inner sep=0] (H) at (0.875,0.375) {};
\draw[very thin, gray, fill=color_0,fill opacity=0.3] (0.000000,0.500000) rectangle (0.250000,0.750000);
% \node[inner sep=0] (I) at (0.125,0.625) {};
\draw[very thin, gray, fill=color_0,fill opacity=0.3] (0.250000,0.500000) rectangle (0.500000,0.750000);
% \node[inner sep=0] (J) at (0.375,0.625) {};
\draw[very thin, gray, fill=color_0,fill opacity=0.3] (0.000000,0.750000) rectangle (0.250000,1.000000);
% \node[inner sep=0] (K) at (0.125,0.875) {};
\draw[very thin, gray, fill=color_0,fill opacity=0.3] (0.250000,0.750000) rectangle (0.500000,1.000000);
% \node[inner sep=0] (L) at (0.375,0.875) {};
\draw[very thin, gray, fill=color_1,fill opacity=0.3] (0.500000,0.500000) rectangle (0.750000,0.750000);
% \node[inner sep=0] (M) at (0.625,0.625) {};
\draw[very thin, gray, fill=color_1,fill opacity=0.3] (0.750000,0.500000) rectangle (1.000000,0.750000);
% \node[inner sep=0] (N) at (0.875,0.625) {};
\draw[very thin, gray, fill=color_1,fill opacity=0.3] (0.500000,0.750000) rectangle (0.750000,1.000000);
% \node[inner sep=0] (O) at (0.625,0.875) {};
\draw[very thin, gray, fill=color_1,fill opacity=0.3] (0.750000,0.750000) rectangle (1.000000,1.000000);
% \node[inner sep=0] (P) at (0.875,0.875) {};
\end{tikzpicture}
        \caption{}
        \label{fig:2d_simple:discretization_oct}
    \end{subfigure}
    \begin{subfigure}{0.7\textwidth}
        \centering
        % cf. https://tex.stackexchange.com/questions/332300/draw-lines-on-top-of-tikz-forest
\forestset{%
    declare keylist register={through},
    through={},
    tracing tree/.style={%
    delay={%
        for #1={%
        if phantom={}{through+/.option=name},
        }
    },
    before drawing tree={%
        tikz+/.wrap pgfmath arg={%
        \foreach \i [count=\j, remember=\i as \k] in {##1} \ifnum\j>1 \draw [densely dashed, ->] (\k.west) -- (\i.west)\fi;
        }{(through)}
    },
    }
}
\tikzset{every label/.style={xshift=-2ex,yshift=-2ex , font=\footnotesize, text=red}}
\providecolor{color_0}{RGB}{255,255,255}
\providecolor{color_1}{RGB}{0, 0, 0}
\begin{forest}
     % tracing tree=tree,
    [11,label={}
        [11,label={}
            [00,circle,draw,fill=color_0,fill opacity=0.4,label={}]
            [00,circle,draw,fill=color_1,fill opacity=0.4,label={}]
            [00,circle,draw,fill=color_0,fill opacity=0.4,label={}]
            [00,circle,draw,fill=color_1,fill opacity=0.4,label={}]
        ]
        [11,label={}
            [00,circle,draw,fill=color_1,fill opacity=0.4,label={}]
            [00,circle,draw,fill=color_1,fill opacity=0.4,label={}]
            [00,circle,draw,fill=color_1,fill opacity=0.4,label={}]
            [00,circle,draw,fill=color_1,fill opacity=0.4,label={}]
        ]
        [11,label={}
            [00,circle,draw,fill=color_0,fill opacity=0.4,label={}]
            [00,circle,draw,fill=color_0,fill opacity=0.4,label={}]
            [00,circle,draw,fill=color_0,fill opacity=0.4,label={}]
            [00,circle,draw,fill=color_0,fill opacity=0.4,label={}]
        ]
        [11,label={}
            [00,circle,draw,fill=color_1,fill opacity=0.4,label={}]
            [00,circle,draw,fill=color_1,fill opacity=0.4,label={}]
            [00,circle,draw,fill=color_1,fill opacity=0.4,label={}]
            [00,circle,draw,fill=color_1,fill opacity=0.4,label={}]
]]]
\end{forest}
        \caption{}
        \label{fig:2d_simple:tree_oct}
    \end{subfigure}\\
    \begin{subfigure}{0.2\textwidth}
        \centering
        \provideboolean{showvalues}
\begin{tikzpicture}
\draw[very thin, white] (0,0) rectangle (1,1);
\providecolor{color_0}{RGB}{255,255,255}
\providecolor{color_1}{RGB}{0, 0, 0}
\draw[very thin, gray, fill=color_0,fill opacity=0.3] (0.000000,0.000000) rectangle (0.250000,0.500000);
\draw[very thin, gray, fill=color_1,fill opacity=0.3] (0.250000,0.000000) rectangle (0.500000,0.500000);
\draw[very thin, gray, fill=color_1,fill opacity=0.3] (0.500000,0.000000) rectangle (1.000000,0.500000);
\draw[very thin, gray, fill=color_0,fill opacity=0.3] (0.000000,0.500000) rectangle (0.500000,1.000000);
\draw[very thin, gray, fill=color_1,fill opacity=0.3] (0.500000,0.500000) rectangle (1.000000,1.000000);
\ifthenelse{\boolean{showvalues}}{
\node[inner sep=0] (A) at (0.125,0.25) {\tiny 1};
\node[inner sep=0] (B) at (0.375,0.25) {\tiny 0};
\node[inner sep=0] (C) at (0.75,0.25) {\footnotesize 0};
\node[inner sep=0] (D) at (0.25,0.75) {\footnotesize 1};
\node[inner sep=0] (E) at (0.75,0.75) {\footnotesize 0};
}{}
\end{tikzpicture}
        \caption{}
        \label{fig:2d_simple:discretization_5}
    \end{subfigure}
    \begin{subfigure}{0.7\textwidth}
        \centering
        \includestandalone[scale=0.5]{gfx/tree_2d_5boxes} \ra{}
        \includestandalone[scale=0.5]{gfx/tree_2d_5boxes_downsplit}
        \caption{}
        \label{fig:2d_simple:tree_5}
    \end{subfigure}\\
    \begin{subfigure}{0.2\textwidth}
        \centering
        \provideboolean{showvalues}
\begin{tikzpicture}
\draw[very thin, white] (0,0) rectangle (1,1);
\providecolor{color_0}{RGB}{255,255,255}
\providecolor{color_1}{RGB}{0, 0, 0}
\draw[very thin, gray, fill=color_0,fill opacity=0.3] (0.000000,0.000000) rectangle (0.250000,0.500000);
\draw[very thin, gray, fill=color_1,fill opacity=0.3] (0.250000,0.000000) rectangle (0.500000,0.500000);
\draw[very thin, gray, fill=color_0,fill opacity=0.3] (0.000000,0.500000) rectangle (0.500000,1.000000);
\draw[very thin, gray, fill=color_1,fill opacity=0.3] (0.500000,0.000000) rectangle (1.000000,1.000000);
\ifthenelse{\boolean{showvalues}}{
\node[inner sep=0] (A) at (0.125,0.25) {\tiny 1};
\node[inner sep=0] (B) at (0.375,0.25) {\tiny 0};
\node[inner sep=0] (C) at (0.25,0.75) {\footnotesize 1};
\node[inner sep=0] (D) at (0.75,0.5) {\footnotesize 0};
}{}
\end{tikzpicture}
        \caption{}
        \label{fig:2d_simple:discretization_4}
    \end{subfigure}
    \begin{subfigure}{0.7\textwidth}
        \centering
        \includestandalone[scale=0.5]{gfx/tree_2d_4boxes}
        \caption{}
        \label{fig:2d_simple:tree_4}
    \end{subfigure}\\
    \caption{
       Two-dimensional example for coarsening and downsplit + coarsening on binary data: 
       The fully-refined octree (\cref{fig:2d_simple:discretization_oct,fig:2d_simple:tree_oct}) contains some redundant information.
       By omnitree coarsening, the tree can be significantly simplified to only five leaf nodes, cf. \cref{fig:2d_simple:discretization_5}, \cref{fig:2d_simple:tree_5} (left).
       Note that octree coarsening alone would not have allowed for any coarsening in the lower left quadrant.
       To further compress the omnittree by fusing the two rightmost quadrants, it is necessary to downsplit the vertical split from the root node \cref{fig:2d_simple:tree_5} (right).
       If the downsplit happens this way, the same-valued leaves become siblings and can be fused with the plain coarsening algorithm, resulting in \cref{fig:2d_simple:discretization_4,fig:2d_simple:tree_4}.
    }
    \label{fig:2d_simple}
\end{figure*}

Downsplit can be considered the inverse to normalization (see \cite[Definition 2.2]{pollingerBeautyAnisotropicMesh2025}):
Normalization absorbs a common child refinement into the parent; downsplit moves a parent refinement down to the children.
The number of leaves in the tree as well as their spatial positions stay the same, but their order in the linearized omnitree may change.

A downsplit rotation can be implemented by reusing the refinement algorithm from \cite{pollingerBeautyAnisotropicMesh2025} as well as the group resolution described in \cref{sec:coarsening} for coarsening.
Conceptually, $v$ is coarsened but in the same step its children are refined in such a way that their descendants remain unaffected (apart from potential reordering).
The only difference to coarsening is that for downsplit, the children are not necessarily leaves and the special case of removing all splits cannot occur.
In the mapping, c.f. \cref{subsec:mapping}, the only tree node that changes structure is the $v$ that is being split down.
If we denote $v$'s index $i$, $v'$'s index $i'$, and $u'_r$'s index $i_r$ (for each $r \in \{0,\ldots, 2^{\abs{S_2}} \}$), then $M[i]$ will consist of the union of $\{i\}$ and $i_r$.

To illustrate the potential benefits of downsplit, \cref{fig:2d_simple} shows different omnitree representations of the same two-dimensional binary data. 
\Cref{fig:2d_simple:discretization_oct} is a regular octree representation with 16 data coefficients.
If we were using only octree compression here, we could fuse into a single leaf all quadrants except the lower left one, and store seven coefficients in total.
Omnitree coarsening as outlined in \cref{sec:coarsening} allows to coarsen the lower left quadrant in one dimension, resulting in five coefficients as can be seen in \cref{fig:2d_simple:discretization_5}.
The last possible optimization for this example is only enabled in combination with downsplit:
\Cref{fig:2d_simple:tree_5} shows how the tree is transformed, while the spatial discretization stays the same.
After the downsplit transformation, the right half can become a single leaf by applying the plain coarsening again.

%TODO downsplit complexity
%TODO should it contain the scanning complexity to find all grandchildren? worst or expected case?

To make use of both coarsening and downsplit, one can apply them in alternation as more and more leaves fuse together. 
But how do we know where coarsening is desirable, and what heuristics can we use to know where and which dimensions to split down?
As the next section will show, wavelet representations can supply answers to both questions.

\section{Haar Wavelets on Omnitrees}\label{sec:wavelets}

% "There is a simple and constructive way of generating tree approximations of a given function f ∈ Lp( ) by thresholding its wavelet coefficients."~\cite{cohenTreeApproximationOptimal2001}
Wavelet function spaces~\cite{mallatTheoryMultiresolutionSignal1989} have been shown to be ideally suited for quad- and octree structured data~\cite{cohenTreeApproximationOptimal2001}, and are also amenable to anisotropic functions~\cite{bhatiaAMMAdaptiveMultilinear2022}.
These \enquote{wavelets on trees} are not to be confused with \enquote{wavelet trees}, which are concerned with string storage.
They also should not be confused with hyperbolic / sparse grid wavelet spaces~\cite{bungartzSparseGrids2004a, hemkerSparsegridFinitevolumeMultigrid1995a, wangSparseGridDiscontinuous2016,schaferHyperbolicWaveletAnalysis2021}, where the hierarchical structure is characterized by intentional spatial overlap of basis functions at different mixed resolutions.

%general wavelet + haar introduction, different conventions
This section defines Haar wavelets on the anisotropic refinement structure of omnitrees and derives the coefficient transformations under coarsening and downsplit.
For simplicity, we consider only piecewise constant functions $f \in L^2([0,1]^d)$ discretized on an omnitree,
\begin{equation}
  f = \sum_{v :\ \vec{b}(v) = \vec{0}} s^v \cdot \phi^v , \qquad
  \phi^v(\vec{x}) = \begin{cases}
                1 & \text{if } \vec{x} \in \text{rect}(v), \\
                0 & \text{else}.
            \end{cases}
\end{equation}
Here, $v$ is a leaf node corresponding to rectangle $\text{rect}(v)$; $s^v$ is the data coefficient at this node; and $\phi^v$ is a nodal basis function for the rectangle.
We call storing $f$ as a combination of the omnitree and the coefficients $s^v$ the \emph{nodal} representation.

\subsection{Tensor-product Haar basis}\label{sec:haar_basis}
Equivalently to the nodal representation, $f$ can be stored \emph{hierachically} using a multiresolution wavelet approach.
We consider the simplest wavelets, Haar wavelets, which naturally represent piecewise constant functions.
In one dimension, the scaling and detail functions associated to an interval, $I=(x^-, x^+)$, are of the form
\begin{equation}
    \hat\phi^I(x) = \begin{cases}
        1 & \text{if } x \in I \\
        0 & \text{else}
    \end{cases}\qquad
    \hat\psi^I = \begin{cases}
        1 & \text{if } x \in \left (x^-, \frac{x^- + x^+}{2} \right) \\
        -1 & \text{if } x \in \left (\frac{x^- + x^+}{2}, x^+ \right) \\
        0 & \text{else}
    \end{cases}
\end{equation}

The extension to multiple dimensions involves a tensor product of the one-dimensional basis.
Within a hierarchy of rectangles generated by an omnitree, several wavelet functions $\psi^v_\tau$ are defined at a node $v$:
\begin{equation}
    \psi^v_\tau(\vec{x}) = \prod_{j \not \in \tau} \hat\phi^{I_j}(x_j) \prod_{j \in \tau} \hat\psi^{I_j}(x_j), \quad \tau \in 2^{\sigma(v)},
\end{equation}
where $\text{rect}(v)$ is a product of the intervals $I_j, \, j \in \sigma(v)$.
Here, the index is over all subsets, $\tau \in 2^{\sigma(v)}$ (i.e., $\tau \subset \sigma(v)$), because a wavelet is only needed in the split dimensions $\sigma(v)$.
The functions $\{\psi^v_\emptyset\}$ are scaling functions, equal to the nodal basis functions $\{\phi^v\}$.
The functions $\{\psi^v_\tau,\, \tau \neq \emptyset\}$, are the detail functions.
The  scaling and wavelet functions form a hierarchical basis containing the same information as the nodal basis formulation, as outlined in the following.

Haar wavelets associated to omnitree nodes $v$ are orthogonal
\begin{equation}\label{eq:wavelet_orth}
    \langle \psi^{v}_{\tau}, \psi^{v'}_{\tau'} \rangle = \delta_{v, v'} \delta_{\tau, \tau'}  .
\end{equation}
Intuitively, this holds because, within a node, the product is odd unless $\tau=\tau'$, and, between nodes, the supported intervals are either disjoint or dyadically nested.
If nested, each smaller (descendant) wavelet's support \enquote{sees} a constant contribution from any ancestor wavelet and multiplies a smaller-scale $0$-mean variation.

Due to orthogonality of the wavelet basis, $f$'s wavelet coefficients $w^v_\tau$ are equivalent to the scalar product of $f$ with the wavelet itself
\begin{equation}
    % f = \sum_{i : \vec{b}(v^i) \neq \vec{0}} \sum_{t \in T(\sigma(v^i))} w_t^i \cdot \psi^i
    w_t^i = \langle f, \psi_t^{({v^{i}})}\rangle, 
\end{equation}
which is computed implicitly by the forward wavelet transform.
(For the multidimensional setting, we use $(w_\emptyset, w_{\{j\}}, \ldots)$ in place of $(s, d)$ common in the wavelet literature.)
See Daubechies~\cite{daubechiesTenLecturesWavelets1992} and Mallat~\cite{mallatTheoryMultiresolutionSignal1989} for detailed mathematical properties of wavelet transforms.

Relevant to our purpose is the fact that the forward wavelet transform can be computed in a bottom-up fashion in linear time.
Let $v$ be an internal node with splitting label $\vec{b}(v)$, refined dimensions $\sigma(v)$, and $2^{\abs{\sigma(v)}}$ child nodes $\{c\}$.
In the base case, the children are leaves.
Associated to their rectangles are nodal coefficients, which may be interpreted as scaling values $w^c_\emptyset$.
We assume that the Z order is used both for ordering the children and their scaling coefficients, which we gather into a small vector, $\vec{s}^v \gets \{w^c_\emptyset\}$.
The wavelet coefficients $\vec{w}^v$ are arranged in a compatible order.
Then, the wavelet coefficients are given by multiplication with a local transformation matrix $H$ at the parent rectangle $\text{rect}(v)$:
\begin{equation}
    \vec{w}^v = H_{\abs{\sigma(v)}} \cdot \vec{s}^v \ .
\end{equation}
For Haar wavelets, the transformations correspond to left-multiplication by the Hadamard matrices
\begin{equation}\label{eq:hier_matrix}
    H_{\abs{\sigma(v)}} = \frac{1}{2^{\abs{\sigma(v)}}} \bigotimes_{j \in \sigma(v)} \begin{pmatrix} 1 & 1 \\ 1 & -1 \end{pmatrix},
    \quad
    R_{\abs{\sigma(v)}} = \bigotimes_{j \in \sigma(v)} \begin{pmatrix} 1 & 1 \\ 1 & -1 \end{pmatrix},
\end{equation}
where $H_{\abs{\sigma(v)}}$ is the hierarchization (decomposition) matrix and $R_{\abs{\sigma(v)}}$ is the nodalization (reconstruction) matrix, satisfying $R_{\abs{\sigma(v)}} H_{\abs{\sigma(v)}} = I$.
The result includes the scaling coefficient $w^v_\emptyset$, which is consumed at the next higher level to compute the wavelet coefficients for the parent of $v$.

Once the local transformations have been applied at all scales, the root scaling coefficient holds the mean value over $(0,1)^d$, and all other information is encoded into wavelet coefficients at non-leaf nodes.
This yields a unique representation of $f$ in the wavelet basis:
\begin{equation}
 f = w^{\text{root}}_\emptyset + \sum_{v: \text{ nonleaf}} \; \sum_{\emptyset \neq \tau \subset \sigma(v)} w^v_\tau \cdot \psi^v_\tau
\end{equation}
Since $\psi^v_\tau \in \{0, 1, -1\}$, the triangle inequality on $\abs{f}$ yields the following bound:
\begin{equation}
    \norm{f}_1 \; \leq \; \abs{w^{\text{root}}_\emptyset} \abs{\Omega} + \sum_{v: \text{ nonleaf}} \; \sum_{\emptyset \neq \tau \subset \sigma(v)} \abs{w^v_\tau} \abs{\text{rect}(v)}
\end{equation}

For any $\tau \neq \emptyset$, $w_\tau$ is the \emph{detail coefficient} measuring variation along exactly the dimensions in $\tau$.
Conversely, if a wavelet coefficient $w^v_\tau$ with $\abs{\tau} = 1$ is zero, it means that change along the respective dimension is not visible within $\text{rect}(v)$ at this particular scale.
From this representation, the original nodal coefficients can be reconstructed by applying the inverse transformation successively from larger to finer scales, i.e. from the root down to the leaf nodes of the tree.

% The inverse transform (nodalization) recovers the nodal values:
% \begin{equation}\label{eq:nodalize}
%     f_e = \sum_{T \subseteq \sigma(v)} w_t^{(v)} \cdot \prod_{j \in T} (-1)^{e_j}.
% \end{equation}

% When the refined dimensions are enumerated as $\sigma(v) = \{j_1, \ldots, j_{\abs{\sigma(v)}}\}$, 

The small multiplications by $H$ and $R$ can be applied to an omnitree discretization (with $N$ leaf nodes) by using data stacks, leading to a maximum transform complexity of $\complexity{(2^d)^2 \cdot \frac{N}{2^d}} = \complexity{2^d N}$ -- linear in the number of coefficients $N$.

\begin{figure}[!htbp]
    \centering
    \begin{subfigure}{0.99\textwidth}
        \centering
        \providecolor{cellneg}{RGB}{200,30,30}
\providecolor{cellpos}{RGB}{30,30,30}
\begin{tikzpicture}[3d view={55}{30}]
\providecommand{\providetikzstyle}[2]{%
    \pgfkeysifdefined{/tikz/#1/.@cmd}{}{\tikzset{#1/.style={#2}}}%
}
\providetikzstyle{connector}{thin, black!25}
\providetikzstyle{leaf}{green, fill=green!20}
\newcommand{\namedrect}[4]{%% Named rectangle: #1=prefix, #2=origin(SW), #3=width, #4=height
\path #2 coordinate (#1-sw)
        ++(#3,0) coordinate (#1-se)
        ++(0,#4) coordinate (#1-ne)
        ++(-#3,0) coordinate (#1-nw);
\coordinate (#1-mid) at ($(#1-sw)!0.5!(#1-ne)$);
\coordinate (#1-mid-s) at ($(#1-sw)!0.5!(#1-se)$);
\coordinate (#1-mid-n) at ($(#1-nw)!0.5!(#1-ne)$);
\coordinate (#1-mid-w) at ($(#1-sw)!0.5!(#1-nw)$);
\coordinate (#1-mid-e) at ($(#1-se)!0.5!(#1-ne)$);
}
%% wavelet macros: #1=prefix, #2=origin(SW), #3=width, #4=height
%% d^H: neg left, pos right (split in x)
\newcommand{\wavH}[4]{%
\namedrect{#1}{#2}{#3}{#4}%
\fill[cellpos] (#1-sw) -- (#1-mid-s) -- (#1-mid-n) -- (#1-nw) -- cycle;
\fill[cellneg] (#1-mid-s) -- (#1-se) -- (#1-ne) -- (#1-mid-n) -- cycle;
\draw[white, semithick] (#1-mid-s) -- (#1-mid-n);
\draw[semithick] (#1-sw) -- (#1-se) -- (#1-ne) -- (#1-nw) -- cycle;
\node[text=white, font=\Huge\bfseries] at ($(#1-sw)!0.5!(#1-mid-n)$) {$+$};
\node[text=white, font=\Huge\bfseries] at ($(#1-mid-s)!0.5!(#1-ne)$) {$-$};
}
%% d^V: pos bottom, neg top (split in y)
\newcommand{\wavV}[4]{%
\namedrect{#1}{#2}{#3}{#4}%
\fill[cellpos] (#1-sw) -- (#1-se) -- (#1-mid-e) -- (#1-mid-w) -- cycle;
\fill[cellneg] (#1-mid-w) -- (#1-mid-e) -- (#1-ne) -- (#1-nw) -- cycle;
\draw[white, semithick] (#1-mid-w) -- (#1-mid-e);
\draw[semithick] (#1-sw) -- (#1-se) -- (#1-ne) -- (#1-nw) -- cycle;
\node[text=white, font=\Huge\bfseries] at ($(#1-sw)!0.5!(#1-mid-e)$) {$+$};
\node[text=white, font=\Huge\bfseries] at ($(#1-mid-w)!0.5!(#1-ne)$) {$-$};
}
%% d^D: checkerboard (BL=neg, BR=pos, TL=pos, TR=neg)
\newcommand{\wavD}[4]{%
\namedrect{#1}{#2}{#3}{#4}%
\fill[cellpos] (#1-sw) -- (#1-mid-s) -- (#1-mid) -- (#1-mid-w) -- cycle;
\fill[cellneg] (#1-mid-s) -- (#1-se) -- (#1-mid-e) -- (#1-mid) -- cycle;
\fill[cellneg] (#1-mid-w) -- (#1-mid) -- (#1-mid-n) -- (#1-nw) -- cycle;
\fill[cellpos] (#1-mid) -- (#1-mid-e) -- (#1-ne) -- (#1-mid-n) -- cycle;
\draw[white, semithick] (#1-mid-s) -- (#1-mid-n);
\draw[white, semithick] (#1-mid-w) -- (#1-mid-e);
\draw[semithick] (#1-sw) -- (#1-se) -- (#1-ne) -- (#1-nw) -- cycle;
\node[text=white, font=\Huge\bfseries] at ($(#1-sw)!0.5!(#1-mid)$) {$+$};
\node[text=white, font=\Huge\bfseries] at ($(#1-se)!0.5!(#1-mid)$) {$-$};
\node[text=white, font=\Huge\bfseries] at ($(#1-nw)!0.5!(#1-mid)$) {$-$};
\node[text=white, font=\Huge\bfseries] at ($(#1-ne)!0.5!(#1-mid)$) {$+$};
}
\begin{scope}[canvas is xy plane at z=-0,transform shape]
\namedrect{zero}{(0,0)}{4}{4}
\draw[semithick,blue,fill=cellpos] (zero-sw) rectangle (zero-ne);
\node[text=white, font=\Huge\bfseries] at (zero-mid) {$+$};
\end{scope}
\begin{scope}[canvas is xy plane at z=-9,transform shape]
\namedrect{onemain}{(0,0)}{4}{4}
\draw[blue] (onemain-sw) grid[step={pow(2,2-1)}] (onemain-ne);
\draw[semithick,blue] (onemain-sw) coordinate(p1) rectangle (onemain-ne);
\draw[thick,->] (-0.2,-0.2) -- (-0.2,0.3) node[above,font=\Large] {${y}$};
\draw[thick,->] (-0.2,-0.2) -- (0.3,-0.2) node[right,font=\Large] {${x}$};
\wavH{onedH}{(0,5)}{4}{4}
\wavV{onedV}{(-5,0)}{4}{4}
\wavD{onedD}{(-5,5)}{4}{4}
\end{scope}
\begin{scope}[canvas is xy plane at z=-15,transform shape]
\namedrect{twoleftmain}{(-2,-2)}{2}{2}
\draw[semithick,blue] (twoleftmain-sw) rectangle (twoleftmain-ne);
\draw[blue] (twoleftmain-mid-n) -- (twoleftmain-mid-s);
\draw[thick,->] (-2.2,-2.2) -- (-2.2,-1.7) node[above,font=\Large] {${y}$};
\draw[thick,->] (-2.2,-2.2) -- (-1.7,-2.2) node[right,font=\Large] {${x}$};
\wavH{twodH}{(-2,0.5)}{2}{2}
\namedrect{twofront}{(4,-2)}{2}{2}
\draw[leaf] (twofront-sw) rectangle (twofront-ne);
\namedrect{twoback}{(-2,4)}{2}{2}
\draw[leaf] (twoback-sw) rectangle (twoback-ne);
\namedrect{tworight}{(4,4)}{2}{2}
\draw[leaf] (tworight-sw) rectangle (tworight-ne);
\end{scope}
\begin{scope}[canvas is xy plane at z=-21,transform shape]
\namedrect{threeleft}{(-2.5,-2)}{1}{2}
\draw[leaf] (threeleft-sw) rectangle (threeleft-ne);
\namedrect{threeright}{(-0.5,-2)}{1}{2}
\draw[leaf] (threeright-sw) rectangle (threeright-ne);
\end{scope}
% % Level 1 connections
% \draw[connector] (onedV-se) -- (onemain-sw);
% \draw[connector] (onedV-ne) -- (onemain-nw);
% \draw[connector] (onedD-sw) -- (onemain-sw);
% \draw[connector] (onedD-ne) -- (onemain-ne);
% \draw[connector] (onedH-sw) -- (onemain-nw);
% \draw[connector] (onedH-se) -- (onemain-ne);
% % Level 2 connections
% \draw[connector] (twodH-sw) -- (twoleftmain-nw);
% \draw[connector] (twodH-se) -- (twoleftmain-ne);
% Level 0 → Level 1 corners
\draw[connector] (zero-sw) -- (onemain-sw);
\draw[connector] (zero-se) -- (onemain-se);
\draw[connector] (zero-ne) -- (onemain-ne);
\draw[connector] (zero-nw) -- (onemain-nw);
% Level 1 → Level 2
\draw[connector] (onemain-sw) -- (twoleftmain-sw);
\draw[connector] (onemain-mid) -- (twoleftmain-ne);
\draw[connector] (onemain-mid-w) -- (twoleftmain-nw);
\draw[connector] (onemain-mid-s) -- (twoleftmain-se);
\draw[connector] (onemain-mid-s) -- (twofront-sw);
\draw[connector] (onemain-mid-e) -- (twofront-ne);
\draw[connector] (onemain-mid) -- (twofront-nw);
\draw[connector] (onemain-se) -- (twofront-se);
\draw[connector] (onemain-mid-w) -- (twoback-sw);
\draw[connector] (onemain-mid-n) -- (twoback-ne);
\draw[connector] (onemain-nw) -- (twoback-nw);
\draw[connector] (onemain-mid) -- (twoback-se);
\draw[connector] (onemain-mid) -- (tworight-sw);
\draw[connector] (onemain-ne) -- (tworight-ne);
\draw[connector] (onemain-mid-n) -- (tworight-nw);
\draw[connector] (onemain-mid-e) -- (tworight-se);
% Level 2 → Level 3
\draw[connector] (twoleftmain-sw) -- (threeleft-sw);
\draw[connector] (twoleftmain-mid-s) -- (threeleft-se);
\draw[connector] (twoleftmain-mid-n) -- (threeleft-ne);
\draw[connector] (twoleftmain-nw) -- (threeleft-nw);
\draw[connector] (twoleftmain-mid-s) -- (threeright-sw);
\draw[connector] (twoleftmain-se) -- (threeright-se);
\draw[connector] (twoleftmain-ne) -- (threeright-ne);
\draw[connector] (twoleftmain-mid-n) -- (threeright-nw);
%draw level 0 again
\begin{scope}[canvas is xy plane at z=-0,transform shape]
\namedrect{zero}{(0,0)}{4}{4}
\draw[semithick,blue,fill=cellpos] (zero-sw) rectangle (zero-ne);
\node[text=white, font=\Huge\bfseries] at (zero-mid) {$+$};
\end{scope}
\end{tikzpicture}
        \caption{Association of function values to leaf rectangles (green) and wavelet coefficients to nonleaf super-rectangles (red/black). Multiple split dimensions lead to a local wavelet space of tensor product form.}
        \label{fig:wavelet_layered}
    \end{subfigure}
    \begin{subfigure}[b]{0.6\textwidth}
        \centering
        \includestandalone[width=\textwidth]{gfx/wavelet_function_3d}\\
        \caption{Binary function taking on the nodal values \num{0} (blue) or \num{1} (yellow), composed of wavelets with coefficients ($3/8$); ($3/8$, $-1/8$, $-1/8$); ($1/2$).
        }
        \label{fig:wavelet_function_3d}
    \end{subfigure}\hfill
    \begin{subfigure}[b]{0.3\textwidth}
        \centering
        \includestandalone[width=0.7\textwidth]{gfx/tree_2d_5boxes_color}\\[1em]
        \def\smath#1{\text{\scalebox{.5}{$#1$}}}
\pgfplotsset{colormap/viridis}
\def\pmmin{-0.2}
\def\pmmax{1.0}
\newcommand{\mapcolor}[2]{%
\pgfmathsetmacro{\mctmp}{1000*(#2 - \pmmin)/(\pmmax - \pmmin)}%
\pgfplotscolormapdefinemappedcolor{\mctmp}%
\colorlet{#1}{mapped color}%
}
\newcommand{\mapcolordark}[2]{%
\mapcolor{#1@raw}{#2}%
\colorlet{#1}{#1@raw!70!black}%
}
\mapcolor{colzero}{0}
\mapcolor{colone}{1}
\mapcolor{colSt}{0.375}
\mapcolor{colDHt}{0.375}
\mapcolor{colDVt}{-0.125}
\mapcolor{colDDt}{-0.125}
\mapcolor{colDHonet}{0.5}
\definecolor{color_0}{RGB}{8,92,248}
\definecolor{color_1}{RGB}{68,166,39}
\definecolor{color_2}{RGB}{190,196,19}
\definecolor{color_3}{RGB}{253,170,95}
\definecolor{color_4}{RGB}{249,81,168}
\tikzset{void/.style={fill=gray}}
\begin{tikzpicture}[every node/.style={draw,align=center,text height=2ex,text depth=0.5ex,minimum width=4ex, inner sep=0.2ex, fill opacity=0.4, text opacity=1}]
\matrix [draw=none, matrix of nodes,nodes in empty cells] (firstmatrix)
    {
     11&
     10&
    |[fill=color_0]| 00&
    |[fill=color_1]| 00&
    |[fill=color_2]| 00&
    |[fill=color_3]| 00&
    |[fill=color_4]| 00\\
    };
\newif\ifprintmask
\printmasktrue   % or \printmaskfalse
\ifprintmask
% another matrix, with black for masked-out / no leaf / nonzero descriptor
\matrix [draw=none, matrix of nodes,nodes in empty cells] at ($(firstmatrix.south)-(0,0.2)$) [anchor=north] (secondmatrix)
    { |[void]|   &
    |[void]|     &
    |[color=colone]|  1&
    |[color=colzero]| 0&
    |[color=colzero]| 0&
    |[color=colone]|  1&
    |[color=colzero]| 0\\
    };
  % hierarchical
    \matrix [draw=none, matrix of nodes,nodes in empty cells] at ($(secondmatrix.south)-(0,0.2)$) [anchor=north] (thirdmatrix)
    { |[color=colDHt]|   {$\smath{0.375}$}&
    |[color=colDHonet]|  {$\smath{0.5}$}&%$= 0.5$
    |[void]|  &
    |[void]|  &
    |[void]|  &
    |[void]|  &
    |[void]| \\
    |[color=colDVt]| {$\smath{-0.125}$}\\
    |[color=colDDt]| {$\smath{-0.125}$}\\
    };
    \matrix [draw=none, matrix of nodes,nodes in empty cells] at (thirdmatrix.north west) [anchor=north east] (scalingmatrix)
    { |[color=colSt]| $\smath{0.375}$\\
    };
    \fi
\end{tikzpicture}
\end{document}\\[5em]
        \caption{Storing either nodal or hierarchical values alongside the tree descriptor.
        }
        \label{fig:wavelet_descriptor}
    \end{subfigure}
  \caption{
    Illustrations of the equivalence of nodal and wavelet representations of a function on an omnitree.
    Note how the parent nodes in (c) map to the blue rectangles in (a) and leaves map to the green rectangles.}
  \label{figs:wavelet_trees}
\end{figure}
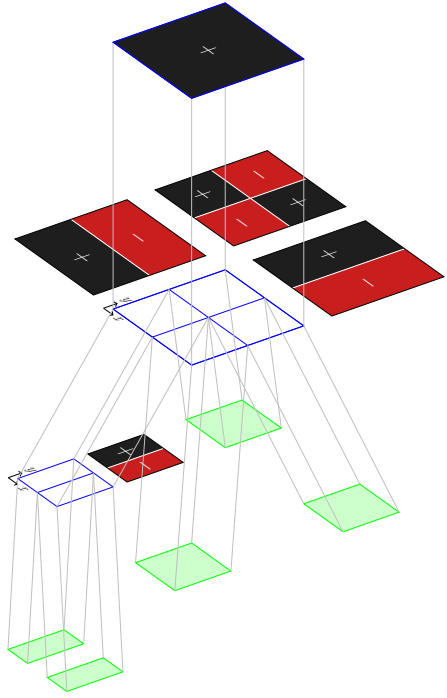
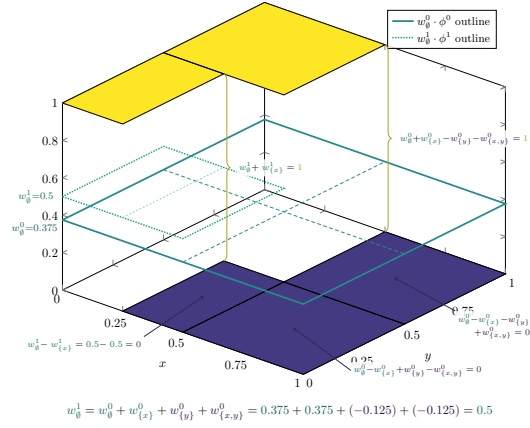
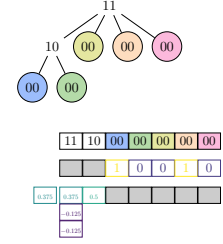

\subsection{Wavelet coefficients under coarsening}\label{sec:coarsening_coefficients}

Wavelet coefficients indicate how much the function changes along any dimension at a given scale.
Some of this information must be preserved when a node is coarsened in a subset of the available dimensions.
When a parent-of-leaves node $v$, i.e. a node with the smallest scales at that spatial position, is coarsened in dimensions $C \subseteq \sigma(v)$, the detail coefficients $w^v_\tau$ with nonempty $\tau \cap C$ are pruned (set to zero), and the remaining coefficients become the coefficient vector of the coarsened node $v'$ with $\sigma(v') = \sigma(v) \setminus C$.
Thus we know that if the wavelet coefficients $w^v_\tau$ for dimensions $j \in \tau$ are (close to) zero, then this node can be well coarsened in dimension $j$.
Upon coarsening, we know that the integral of the function $f$ will be exactly preserved, since the discarded wavelet functions have an integral of exactly zero, due to orthogonality against constant functions (\cref{eq:wavelet_orth}).
Accumulating the weighted magnitudes of discarded detail coefficients, $\sum \abs{w^v_\tau} \abs{\text{rect}(v)}$, provides an upper bound for the overall $L_1$ error incurred in the compression.

% Example in $3$-d (dimensions $\sigma(v)=\{x,y,z\}$):
% Coarsen $C = \{x, y\}$. The dimensions remaining in the new node are $\{z\}$.
% The new wavelet coefficients are $\tau=\emptyset$ and $\tau=\{z\}$.
% In other words, coefficients kept are $\tau \subset \sigma(v) \setminus C$.
% This is equivalent to discarding $\tau \not \subset \sigma(v) \setminus C$,
% which is equivalent to $\tau \cap C$ is nonempty, i.e., $\exists j \in \tau : j \in C$.
 
\subsection{Wavelet coefficients under downsplit}\label{sec:downsplit_coefficients}

Consider the wavelet coefficients at a node $v$, from which dimensions $S_1 \subset \sigma(v)$ are to be downsplit and dimensions $S_2 = \sigma(v) \setminus S_1$ retained.
The node $v$ is replaced by a residual node $v'$ ($\sigma(v') = S_2$) and its intermediate children $\{u_r\}$ ($\sigma(u_r) = S_1$).
The source wavelet coefficients $w^v_\tau, \, \tau \subset \sigma(v)$ must be redistributed accordingly.

The coefficients associated purely to $S_2$ are transferred directly to $v'$:
\begin{equation}\label{eq:new_parent_coeffs}
  w^{v'}_{\tau_2} = w^v_{\tau_2}, \quad \emptyset \neq \tau_2 \subset S_2 \ .
\end{equation}

Meanwhile, the coefficients that involve $S_1$ participate in reconstruction at the intermediate nodes $\{u_r\}$:
\begin{equation}\label{eq:interm_coeffs}
  w^{\{u_r\}}_{\tau_1} = \sum_{\tau_2 \subset S_2} (R_{\abs{S_2}})^{\{r\}}_{\tau_2} \cdot w^v_{\tau_2 \sqcup \tau_1} , \quad \emptyset \neq \tau_1 \subset S_1
\end{equation}
Here, $(R_{\abs{S_2}})$ is a nodalization matrix as in \cref{eq:hier_matrix}, where the Kronecker product is taken over the dimensions in $S_2$.
The rows are indexed by the children of $v'$ in Z order and the columns are indexed by the power set $2^{S_2}$ in a compatible way.
The source coefficients can be interpreted as a tensor with disjoint indices $\tau_1$ and $\tau_2$ referring to distinct modes $S_1$ and $S_2$.

As a matter of accounting, let $k_1 = \abs{S_1}$, $k_2 = \abs{S_2}$, and $k = k_1 + k_2 = \abs{\sigma(v)}$.
The downsplit decouples $2^k - 1$ source coefficients into one parent set of size $(2^{k_2} - 1)$ and $2^{k_2}$ child sets of size $(2^{k_1} - 1)$, then applies identical inverse Haar butterflies to all child sets.
We entirely omit the scaling coefficients $w_{\emptyset}$ for these computations, but in principle they could also be stored for a redundant encoding.

\subsection{Wavelet Compression on Omnitrees}\label{sec:compression}

The coefficient transformations of \cref{sec:coarsening_coefficients,sec:downsplit_coefficients} enable a compression strategy that operates entirely in the wavelet domain:
Coarsening prunes small detail coefficients, and downsplit restructures the tree to expose new coarsening opportunities that multi-dimensional splits would otherwise hide.
Thus, we can transform the nodal input data to wavelet coefficients, and reconstruct the nodal function values only at the end, for output file writing and error evaluation.

Coarsening acts on functions as a linear projection to a reduced set of leaf rectangles.
Downsplit and normalization preserve functions and leaf rectangles, while changing the basis used for the coefficients.
Thus the final set of wavelet coefficients resulting from any sequence of downsplit, coarsening, or normalization passes depends only on the final omnitree.

Selecting the absolute optimal omnitree discretization (in terms of storage) is not computationally feasible at fine resolutions in $3d$.
For now, we consider a heuristic (\cref{sec:compression:downsplit}) that focuses on the parents and grandparents of leaves, as leaves carry the data coefficients and we already see significant savings through this heuristic.

\subsubsection{Coarsening compression}\label{sec:compression:coarsening}
Given a hierarchized omnitree and an error density threshold $\varepsilon \geq 0$, the coarsening pass identifies nodes whose children are leaves and whose detail coefficients are small enough to discard.
The compression scheme schedules such a node $v$ to be coarsened in dimension(s) $C \subseteq \sigma(v)$ if the detail coefficients associated with $C$ satisfy
\begin{equation}\label{eq:coarsening_criterion}
    \abs{ w^v_\tau } \;\leq\; \varepsilon \quad \forall {\tau \cap C \neq \emptyset}. % \cdot \frac{1}{\mathrm{vol}(\mathrm{rect}(v))},
\end{equation}
%where $\mathrm{vol}(\mathrm{rect}(v)) = \prod_{j=1}^{d} 2^{-\ell_j(v)}$ is the volume of the hyperrectangle at node $v$ and $\ell_j(v)$ is the per-dimension level.
When all detail coefficients simultaneously satisfy this criterion, $C$ is set to $\sigma(v)$ and the node is fully coarsened into a leaf.
Otherwise, individual dimensions $j \in \sigma(v)$ whose associated detail coefficients are within the bound can be coarsened, resulting in a partial coarsening that reduces $\abs{\sigma(v)}$ without removing the node entirely.
Coarsening is applied bottom-up from last-level parents (\cref{sec:coarsening}) and repeated until no further coarsening passes are possible.

% The total error incurred by all coarsening passes is bounded by a function of the threshold $\varepsilon$, the dimensionality $d$, the domain volume (we assume unity), and the initial refinement levels $\ell$:
% \begin{subequations}
% \begin{align}
%   \norm{f - f_0}_1 \leq \sum_{v \text{ nonleaf}} (2^{\abs{\sigma(v)}} - 1) \cdot \varepsilon &= (N - 1) \cdot \varepsilon \\
%   &\leq ((2^\ell)^d - 1) \cdot \varepsilon
% \end{align}
% \end{subequations}
% The relations are derived by assuming all parent nodes could be coarsened.
% The number of wavelet coefficients (each contributing error $\varepsilon$) at all parents equals the total number of children minus the total number of parents; the leaves and the root are the only nodes that do not have both parents and children.
% Finally, the number of leafs can be no more than the compounding of $2^d$ children at each level.

The total error incurred by all coarsening passes is bounded by a function of the threshold $\varepsilon$, domain volume, dimension $d$, and source tree height $\ell$:
\begin{equation}
  \norm{\tilde f - f}_1 \; \leq \sum_{v \text{ nonleaf}} (2^{\abs{\sigma(v)}}-1) \cdot \varepsilon \abs{\text{rect}(v)}
  \; \leq \; (2^d-1) \cdot \varepsilon \abs{\Omega} \cdot \ell
\end{equation}
These relations are derived by assuming all parent nodes could be fully coarsened, committing error $\varepsilon$ for each detail coefficient.
Summing parent rectangle volumes level by level, the domain is covered $\ell$ times.

In practice, this \emph{a priori} bound will greatly overestimate the actual $L_1$ error when $\varepsilon > 0$.
A tighter \emph{a posteriori} estimate may be obtained during the compression, by aggregating coefficient magnitudes from the nodes actually selected for coarsening.
This holds also for the downsplit compression discussed in the next section.
When $\varepsilon = 0$, only nodes with exactly zero detail coefficients are coarsened, yielding lossless compression.

\subsubsection{Downsplit-coarsening compression}\label{sec:compression:downsplit}

Omnitree coarsening alone cannot always exploit all redundancy, but some redundancy can be exposed by downsplitting, cf. \cref{fig:2d_simple}.

This motivates an alternating strategy.
After plain coarsening has converged, the algorithm enters the \emph{downsplit-coarsening loop}:
\begin{enumerate}
    \item \textbf{Downsplit.} For each multi-dimensional node $v$ (with $\abs{\sigma(v)} \geq 2$) and at least two leaf children, select the dimension $j^* \in \sigma(v)$ whose pure one-dimensional detail coefficient has the smallest magnitude:
    \begin{equation}\label{eq:dim_selection}
        j^* = \argmin_{j \in \sigma(v)} \abs{ w^v_{\tau_j} },
    \end{equation}
    where $\tau_j = \{j\} \subset \sigma(v)$.
    Downsplit the single dimension $\{j^*\}$ and update all coefficients using \cref{eq:new_parent_coeffs,eq:interm_coeffs}.
    %(and \cref{eq:merged_coeffs} for absorbed children).
    \item \textbf{Coarsen.} Apply the coarsening criterion \cref{eq:coarsening_criterion} to all nodes whose children are all leaves.
    \item \textbf{Normalize.} Resolve any uniqueness violations introduced by the combination of downsplit and coarsening, restoring a canonical omnitree.
\end{enumerate}
These three steps are repeated until no coarsenings occur in a round.

% The downsplit targets are processed bottom-up in order of decreasing depth, so that ancestor--descendant conflicts are avoided within each batch.
The dimension selection heuristic \cref{eq:dim_selection} assumes the dominance of first-order effects, so that when $w^v_{\tau_j}$ is near zero, all intermediates along~$j$ will have negligible detail.
In other words, the dimension with the smallest one-dimensional detail is the one along which the function varies the least, making it the most promising candidate for coarsening after downsplit.

\subsection{Worked Example}\label{sec:worked_example}
Since this section introduces many (partly new) concepts, we illustrate them with a simple example in $2d$.
\Cref{fig:2d_simple:discretization_5} shows a omnitree discretization of a function with $5$ leaf nodes, where the function values are binary when stored in the nodal representation: 0 in the darker and 1 in the lighter rectangles.
In \cref{fig:wavelet_layered}, there is an expanded view of the corresponding tree structure (also shown in \cref{fig:2d_simple:tree_5} left).
Nodal coefficients are associated to the leaf rectangles (green).
Equivalently, wavelet coefficients are associated to the non-leaf rectangles (red/black); note that there are five values to store in either case.

The wavelet coefficients are computed by applying the local transformation matrices $H$ in a bottom-up fashion.
Starting at the leftmost parent node $v^1$, its scaling coefficient $w^1_\emptyset$ and its detail coefficients $w_{\{x\}}$ can be computed by applying the local transformation matrix $H_1$ to the vector of its two children's nodal values
\begin{equation}
    \begin{pmatrix} w^1_\emptyset \\ w_{\{x\}}^1 \end{pmatrix} = \frac{1}{2} \begin{pmatrix} 1 & 1 \\ 1 & -1 \end{pmatrix} \cdot \begin{pmatrix} 1 \\ 0 \end{pmatrix} = \begin{pmatrix} 1/2 \\ 1/2 \end{pmatrix}.
\end{equation}
In line with our definition of the transform matrices, the scaling coefficient is the average of the two nodal values, and the detail coefficient is half of their signed difference.
Using the same procedure, we can compute the coefficients for the root node $v^0$, with its four children $v^1, v^4, v^5, v^6$
\begin{equation}
    \begin{pmatrix} w^0_\emptyset \\ w_{\{x\}}^0 \\ w_{\{y\}}^0 \\ w_{\{x,y\}}^0 \end{pmatrix} = \frac{1}{4} \begin{pmatrix} 1 & 1 & 1 & 1 \\ 1 & -1 & 1 & -1 \\ 1 & 1 & -1 & -1 \\ 1 & -1 & -1 & 1 \end{pmatrix} \cdot \begin{pmatrix} w^1_\emptyset \\ 0 \\ 1 \\ 0 \end{pmatrix} = \begin{pmatrix} 3/8 \\ 3/8 \\ -1/8 \\ -1/8 \end{pmatrix}.
\end{equation}
\Cref{fig:wavelet_function_3d} illustrates how the function can be reconstructed from the wavelet coefficients.
For the data coefficients, we can choose to store either the nodal values at the leaves or the wavelet coefficients at the non-leaves, as shown in \cref{fig:wavelet_descriptor}.
The coarsening criterion \cref{eq:coarsening_criterion} with $\varepsilon = 0$ does not suggest an opporunity for lossless compression, since all detail coefficients are nonzero.
However, if we apply the downsplit-coarsening loop, the root node $v^0$ splits down the $y$-dimension, since the one-dimensional detail coefficient $w_{\{y\}}^0$ has the smallest magnitude.
The resulting tree structure is shown in \cref{fig:2d_simple:tree_5} right.
The root node keeps the coefficient $w_{\{x\}}$, while the details $w_{\{y\}}$ and $w_{\{x,y\}}$ are used to compute the coefficients of its new, intermediate children
\begin{equation}
    \begin{pmatrix} w_{\{y\}}^{u_1} \\ w_{\{y\}}^{u_2} \end{pmatrix} = \begin{pmatrix} {w'}_{\{y\}}^{1} \\ {w'}_{\{y\}}^{6} \end{pmatrix} = \begin{pmatrix} 1 & 1 \\ 1 & -1 \end{pmatrix} \cdot \begin{pmatrix} w_{\{y\}}^0 \\ w_{\{x,y\}}^0 \end{pmatrix} = \begin{pmatrix} -1/4 \\ 0 \end{pmatrix}.
\end{equation}
Now, the intermediate node $u_2$ has zero detail coefficient and can be coarsened.
This removes the leaf nodes $v'^7, v'^8$ by making $v'^6$ a leaf and reduces the total number of coefficients to store from $5$ to $4$ by dropping the ${w'}_{\{y\}}^{6} = 0$ coefficient, with discretization and tree structure as in \cref{fig:2d_simple:discretization_4,fig:2d_simple:tree_4}.
From the remaining wavelet coefficients $(3/8), (3/8), (-1/4), (1/2)$, one can perfectly reconstruct the corresponding binary nodal coefficients $1, 0, 1, 0$ by multiplying with $R_1$ starting at the top of the tree.

\section{Experimental Evaluation and Comparison to OpenVDB}\label{sec:experiment}

In this section we observe the empirical relationship between data resolution and storage size when representing $3$-d data in wavelet form on omnitrees.
The task is to minimize the storage size for a given resolution.
Our experimental methodology varies the resolution in the following ways:
For our evaluation of lossless compression, Boolean density distributions, $f_\ell(\vec{x}) \in \{0,1\}$, are given, pre-voxelized into cubes of a target width $2^{-\ell}$, and we observe the sizes of omnitrees required to represent these functions exactly.
For our evaluation of lossy compression, a real density distribution $f(\vec{x}) \in [0,1] \subset \mathbb{R}$ is given, and coarsened omnitrees are constructed to approximate $f$ to various tolerances $\varepsilon > 0$.
The storage size is measured both as the number of numeric function values stored and as the total number of bytes, including data structure overheads.

For a baseline comparison, we also observe the storage requirement of OpenVDB~\cite{musethVDBHighresolutionSparse2013}, a state-of-the-art hierarchical volumetric data structure, as introduced in \cref{sec:related_work}.
To ensure a direct comparison, the discretized functions are first obtained in the format of OpenVDB.
The storage requirement of OpenVDB is counted as the number of leaf voxels plus the number of active tiles.
The VDB structure is then unfurled to an equivalent octree, stored in the omnitree format.
After applying a plain coarsening pass as described in \cref{sec:compression:coarsening}, we count the number of wavelet coefficients.
Continuing from this result, we further apply the downsplit-coarsening loop (cf. \cref{sec:compression:downsplit}), and count again.
Thus for each resolution three measurements are made---one for OpenVDB, one for omnitrees without downsplit, and one with.

In addition, we compare the file sizes with OpenVDB data stored in a \blosc{}-compressed~\cite{blosc} binary file (this is default for OpenVDB).
For omnitrees, we reconstruct the nodal values and write them to a binary file using \texttt{numpy}.
We consider the raw storage size of the tree descriptor and the data coefficients stored separately, as well as different combinations of raw and \blosc{}-compressed files.

\begin{figure}[!htbp]
    \centering
    \includestandalone[mode=tex,width=0.3\textwidth]{gfx/animated_vdb_cat}
    \caption{Cat object (\href{https://web.archive.org/web/20250509051543/https://ten-thousand-models.appspot.com/detail.html?file_id=100349}{Thingi 34965}) at various resolution levels ($\ell = {2}$ to $\ell = {7}$).
    Animation available with \href{https://tex.stackexchange.com/questions/235139/using-the-animate-package-without-adobe}{various pdf readers}. %; for a single image, the highest resolution $\ell = {7}$ is shown.
    Rendered in Blender~\cite{blender} from VDB file.
    }
    \label{fig:vdb_blender:cat}
\end{figure}
% \todo{Caption says ``is shown..'' is something missing?} -> no, just a typo ;) I've added ell as an overlay anyways now

\subsection{Binary function and lossless compression: Comparison on \num{4166} thingies}\label{sec:results:thingies}

Like in previous work~\cite{pollingerBeautyAnisotropicMesh2025}, we consider the \thingitenk{} data set of surface models curated from a database of printable three-dimensional objects~\cite{zhouThingi10KDataset100002016}.
We filter this data set to the \num{4166} models which contain up to \num{10000} vertices, and which are watertight, not self-intersecting, and solid.
The model coordinates are shifted and scaled to fit in the unit cube.
To obtain a voxel representation at a given level of detail $\ell$, the unit cube is subdivided into $2^\ell$ voxels in each dimension,
and a function $f(\vec{x})$ is defined as $1$ over every voxel whose midpoint is inside the object and $0$ elsewhere.
This setup is more representative of practical numerical approaches such as level sets, in contrast to our previous investigation that allowed almost arbitrarily small refinements~\cite{pollingerBeautyAnisotropicMesh2025}.
\Cref{fig:vdb_blender:cat} shows an example of a representative object, which was also the object of our previous study, at various resolution levels $\ell$.

The samples are used to initialize a VDB \texttt{BoolGrid} structure, using the same voxel size $2^{-\ell}$.
OpenVDB's \texttt{prune()} function is called to condense the hierarchy.
The coefficients are compounded in VDB's hierarchical storage where there is no small-scale change.
After storing this baseline OpenVDB file, we apply the omnitree plain and downsplit coarsening algorithms as described above, setting the threshold as $\varepsilon = 0$ so that only lossless compression takes place.
This allows us to compare storage between OpenVDB and omnitrees for the exact same functions.
Importantly, lossless coarsening ensures that the same binary function is maintained.
This is true, not only in the wavelet basis, but also in the nodal basis, whose binary coefficients are the ones counted in the following.

\begin{figure}[!ht]
    \centering
    \includestandalone[width=0.99\textwidth]{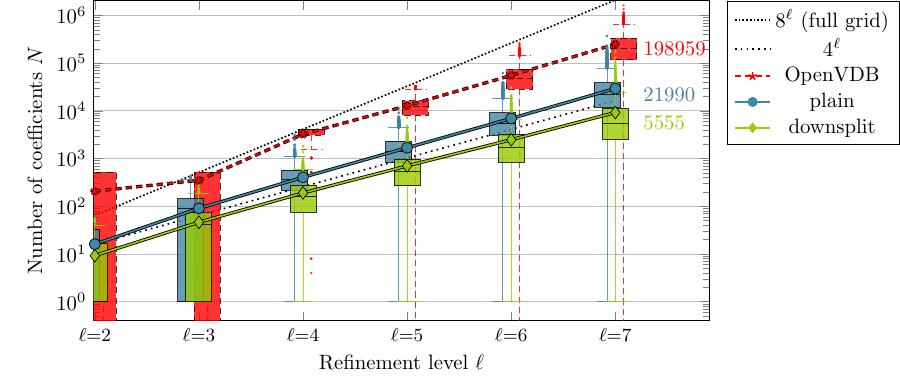}
    \caption{Boxplots and lineplots showing the number of stored binary coefficients to represent the \num{4166} thingies with varying levels of resolution.
    The black dotted line denotes the number of coefficients that would be stored on a uniformly resolved full grid.
    For the boxplots, the box boundaries denote the q25 and q75 quartiles and the line across the box denotes the median.
    The line plots show the mean values, which are consistently slightly higher than the median values.
    At very coarse resolutions, OpenVDB even stores the same number or more coefficients as the full grid, but starting from $\ell=5$, one can clearly see savings.
    Omnitrees (both with only coarsening and downsplit) reach this asymptotic region much earlier, since they store no intermediate hierarchical data and can take advantage of the anisotropy inherent in the three-dimensional objects.
    Median values at $\ell=7$ are annotated in the figure.
    The mean and median difference from OpenVDB to downsplit omnitree at $\ell=7$ amount to \qty[round-precision=2]{96.3002055}{\percent} and \qty[round-precision=2]{97.2079675}{\percent}, respectively.
    Median numbers for $\ell=7$ are printed to the right.
    }
    \label{fig:coeff_boxplot}
\end{figure}

\Cref{fig:coeff_boxplot} shows the number of binary coefficients stored by the three methods.
Plotted are the mean (line) and median, q25 and q75 quartile, and extremal statistics (boxplot) over the \num{4166} objects.
Displayed for reference are slopes of $(4^\ell)$ and $(8^\ell)$ corresponding to $2$-d surface- and $3$-d volume-complexities in the quadtree/octree complexity theorem~\cite{sametHierarchicalDataStructures1988}.

We briefly comment on the variances appearing at the coarsest resolutions, $\ell=2,3$.
Because voxels are initialized by sampling at midpoints, on very coarse grids the interiors of some objects are not sampled at all and appear to be empty.
At $\ell=3$, a large fraction and, at $\ell=2$, a majority of the objects create this effect.
In this situation, our measurements count OpenVDB as storing no values, leading to zero coefficients in \cref{fig:coeff_boxplot}, outside the range of the logarithmic scale.
Omnitrees are counted as storing one value in the same case.
Due to this sampling artifact, the median, lower quartile, and lower whisker may coincide at $\ell=2,3$.

Whether inspecting the mean values or the median values, the overall trends are the same.
The discussion will focus on the median values unless noted otherwise.
(This is for easier comparability with previous work.)
As such, for $\ell > 3$, it is seen that the plain omnitree method uses on average $\qty{8.75}{\xtimes}$ fewer coefficients as OpenVDB.
With downsplit omnitrees, this number is further reduced by at least \qty[round-precision=2]{2.275}{\xtimes} ($\ell=4$) and by up to \qty[round-precision=2]{3.95859585958596}{\xtimes} ($\ell = 7$).

We calculate the empirical approximation rates $r_a$ of the three methods to analyze their asymptotics:
\begin{equation}
   r_a(\ell) = \log_{10}(N_{\ell})-\log_{10}(N_{\ell-1}) , 
\end{equation}
where $N_\ell$ is the number of coefficients at resolution level $\ell$.
Visually, in \cref{fig:coeff_boxplot}, $r_a$ is the slope between $(\ell-1)$ and $\ell$.
Flatter slope means that more coefficients can be pruned at every scale while representing the same data.
The quadtree approximation theorem predicts a theoretical rate of $r_a^* = \log_{10}(2^2) \approx$ \num[round-precision=2]{0.602059991327962}.
We calculate $r_a^\text{VDB}(7) =$ \num[round-precision=2]{0.625646967269006} for OpenVDB and $r_a^\text{plain}(7) =$ \num[round-precision=2]{0.611160511971869} for plain omnitrees, both hovering near the predicted value.
Downsplit omnitrees, however, undercut the prediction at $r_a^\text{downsplit}(7) =$ \num[round-precision=2]{0.514235141898613}.
% For $\ell = 7$, OpenVDB's approximation rate amounts to \num[round-precision=2]{0.625646967269006}, very close to the plain omnitree at \num[round-precision=2]{0.611160511971869}.
% This corresponds well to the expected rate \num[round-precision=2]{0.602059991327962} according to the quadtree approximation theorem.
% However, the approximation rate is reduced to \num[round-precision=2]{0.514235141898613} when using downsplit, significantly undercutting the expectation.
This shows that, on the task of lossless compression, omnitrees asymptotically outperform isotropic data structures like octrees and OpenVDB, and that downsplitting anisotropy is essential to do so.

In absolute terms, there is no single object for which omnitrees require more coefficients than OpenVDB.
The ratio of omnitree coefficients to OpenVDB coefficients for any given object ranges from at most $1$ (e.g. for a perfect cube) down to $\frac{1}{5472}$ at $\ell = 7$ (e.g. for \href{https://web.archive.org/web/20250616141737/https://ten-thousand-models.appspot.com/detail.html?file_id=95807}{Thingi 95807}, a flat cuboid object).

\begin{figure}[!ht]
    \centering
    \includestandalone[width=0.99\textwidth]{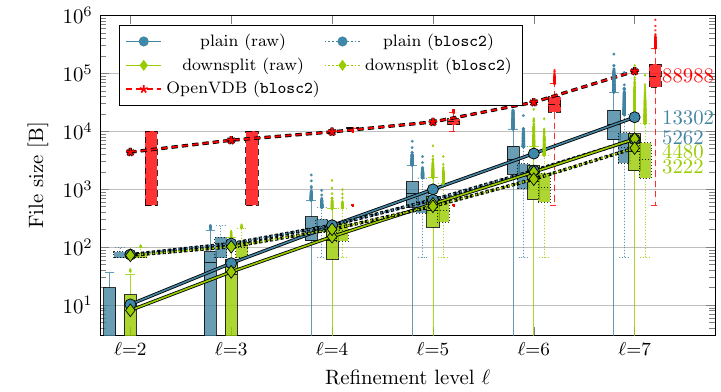} %todo separate median numbers to the right
    \caption{
        Boxplots and lineplots showing the the file size required to store the \num{4166} thingies with varying levels of resolution.
        Colors and styles are used analogously to \cref{fig:coeff_boxplot}, with the addition of dotted properties for \blosc{}-compressed omnitree storage (with compressed descriptor and raw binary data coefficients).
        While \blosc{} compression reduces the file size for all omnitree variants, the effect is more pronounced for the plain coarsening than for downsplit, which already reduces the number of coefficients to store.
    }%TODO relative difference between blosc and raw
    \label{fig:storage_boxplot}
\end{figure}

Next we compare file sizes.
Note that OpenVDB uses \blosc{}~\cite{blosc} compression by default, and we have kept this behavior enabled in our experiments.
In this particular scenario, we are storing only binary numbers as function data (i.e., one bit per leaf as data coefficients), such that the descriptor (i.e. three bits for all tree nodes) dominates the overall storage for omnitrees.
We noticed that for the \thingitenk{} omnitree compression there is rarely a benefit of using \blosc{} compression on the coefficients data for fine resolutions, which is consistent with the high (and increasing) information density observed in \cite[Fig. 7(b)]{pollingerBeautyAnisotropicMesh2025}.
By contrast, the descriptor data is usually well compressible with \blosc{}. 
Accordingly, we consider five cases in \cref{fig:storage_boxplot}:
Each of the omnitree schemes has a variant with raw storage of both binary arrays, and one variant where the coefficients are stored in raw format and the descriptor is compressed with \blosc{}.
Again, downsplit achieves the smallest overheads, but benefits less from compression compared to the plain coarsening.
Using storage instead of coefficients in computing the approximation rate, the OpenVDB solution performs favorable with a rate of \num[round-precision=2]{0.493246428658534}, clearly outperforming the plain and downsplit omnitrees stored raw at \num{0.611966773122632} and \num[round-precision=2]{0.514447441784292}, respectively, while starting at a much larger footprint to begin with.
Using \blosc{} for the omnitree as well, the rates are virtually the same as for OpenVDB, at \num[round-precision=2]{0.499386790847176} and \num[round-precision=2]{0.474769170538491}, with a lower value to start with.
At $\ell = 7$, the median storage size of OpenVDB and omnitree differs by factors between \qty{6.68982107953691}{\xtimes} (for plain coarsening and raw storage) and \qty{27.6145849495733}{\xtimes} (for downsplit coarsening and \blosc{} storage).

% In fact, for the finest resolutions, the compressed plain files achieve basically the same mean size as the uncompressed downsplit-generated files.
% One can speculate that the compression achieved by \blosc{} removes some of the structural redundancy that is also exploited by downsplitting the omnitree.

\subsection{Continuous function and lossy compression: Comparison on Disney cloud}\label{sec:results:cloud}

\begin{figure}[!htbp]
    \centering
    \providecommand{\lcroppts}{500pt}
    \providecommand{\bcroppts}{70pt}
    \providecommand{\rcroppts}{400pt}
    \providecommand{\tcroppts}{100pt}
    \begin{subfigure}[t]{0.45\textwidth}
    \includegraphics[width=\textwidth,trim=500pt 70pt 400pt 100pt, clip]{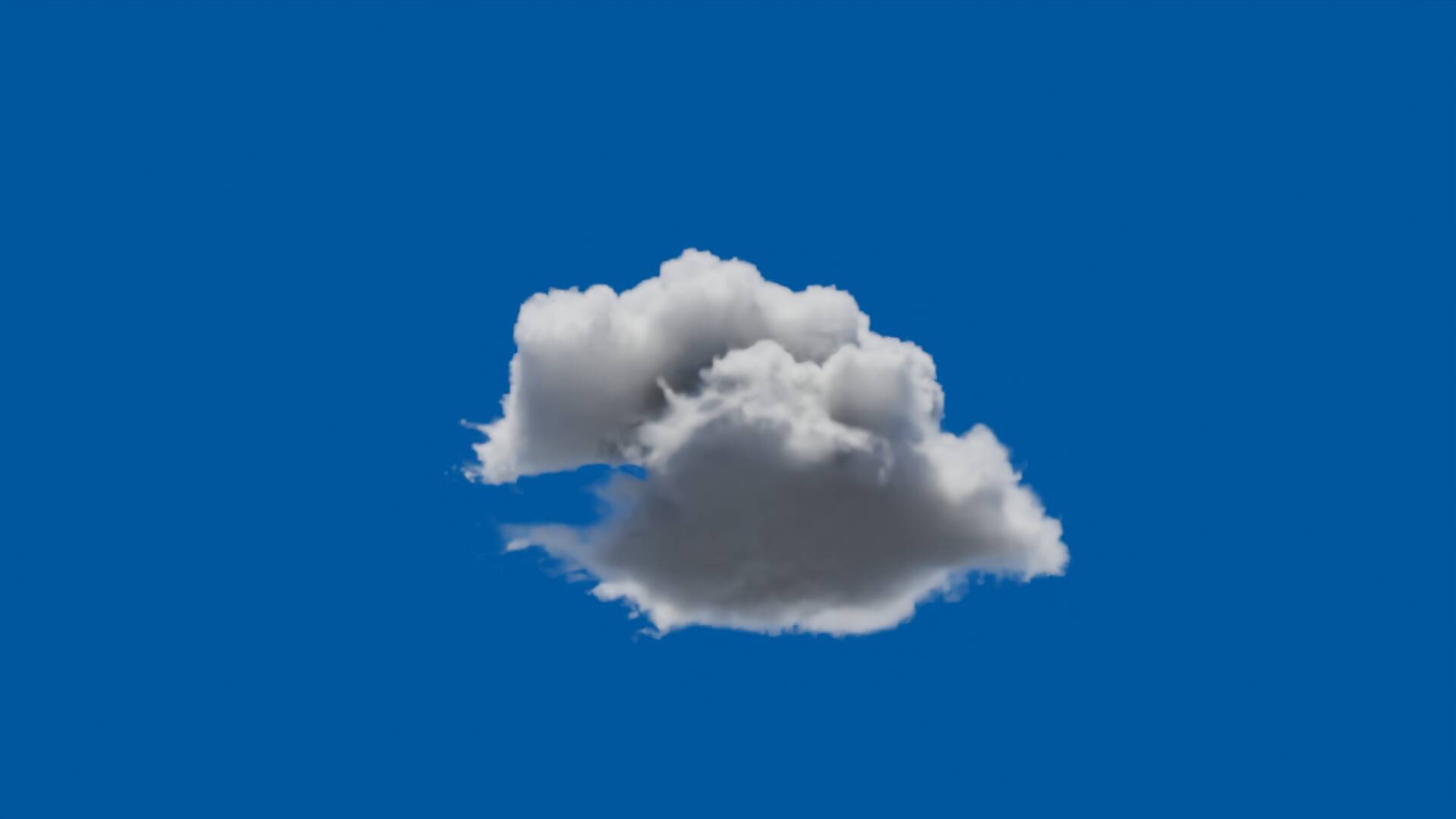}
        \caption{Original sixteenth cloud from WDAS cloud data set containing \num{415642} values.}
        \label{fig:cloud:original}
    \end{subfigure}\\
    \begin{subfigure}[t]{0.49\textwidth}
        \hspace*{-0.5em}
    \providecommand{\cvariant}{can}
    \includestandalone[mode=tex,width=\textwidth]{gfx/animated_cloud}
        \caption{Lossy cloud produced by plain omnitree compression.}
        \label{fig:cloud:lossy:can}
    \end{subfigure}\hfill
    \begin{subfigure}[t]{0.49\textwidth}
        \hspace*{-0.5em}
    \providecommand{\cvariant}{ds}
    \includestandalone[mode=tex,width=\textwidth]{gfx/animated_cloud}
        \caption{Lossy cloud produced by downsplit omnitree compression.}
        \label{fig:cloud:lossy:ds}
    \end{subfigure}
    \caption{
       Disney cloud (one sixteenth) in its original and lossily compressed versions.
       Animations available for $ \varepsilon = \numrange{e-2}{1} $ with \href{https://tex.stackexchange.com/questions/235139/using-the-animate-package-without-adobe}{various pdf readers}.
       Up to $\varepsilon = \num{e-1}$, the visual quality is almost unaffected, while the number of coefficients is already significantly reduced.
       For higher thresholds, the mass is conserved but distributed to larger and larger cuboids, ultimately leading to a single-cuboid low-density \enquote{block}.
       Rendered in Blender~\cite{blender} from VDB file.
    }
    \label{fig:cloud:vdb_blender}
\end{figure}

Walt Disney Animation Studios provides a large volumetric cloud model~\cite{WaltDisneyAnimation} under a Creative Commons license, downloadable in OpenVDB format.
We select the one-sixteenth downsampled version, also provided.
The contained floating-point density distribution $f^*$ directly serves as the baseline data in our omnitree workflow, after zero-padding.
The active extent of the input data is originally $(126, 86, 154)$, facilitated by VDB's top-level sparse map allowing domains of arbitrary bounds.
A corresponding omnitree is constructed with initial resolution $\vec{\ell_0} = (7,7,8)$ from a zero-pad extended voxel grid of extent $(128, 128, 256)$.

Our experiments in this section show the relation between storage size and approximation error under lossy wavelet compression.
These are bounded by the results for lossless compression, which are also shown for reference.
The prescribed lossy wavelet thresholds range from $\varepsilon =$ \num{e-6} to \num{1}.

\Cref{fig:cloud:vdb_blender} (animated in the electronic version) provides a visual understanding of the approximation quality for different thresholds.
In the judgment of the authors, only slight variations in the fine, near-zero density cloud structures are detectable, with almost no other noticeable change up to $\varepsilon = \num{e-1}$.
The cloud begins shifting into blocky cuboids for $\varepsilon = \num{3.3e-1}$.
In this range, though mass is conserved, it is being distributed across larger and larger extents that begin to register visually as the threshold increases.
There is virtually no visual difference between plain and downsplit coarsening.

% \todo{Legend for plain vs downsplit in error plot} -> added
\begin{figure*}[!htbp]
    \centering
    \begin{subfigure}[t]{0.75\textwidth}
    \def\axisdefaultwidth{12cm}
\def\axisdefaultheight{7cm}

\begin{tikzpicture}
  \pgfplotstableread[col sep=comma]{cloud/cloud_wavelet_stats.csv}\cloudtable
\pgfplotstablegetelem{0}{L1_sol}\of\cloudtable
\edef\lonesol{\pgfplotsretval}
\pgfplotstablegetelem{0}{L2_sol}\of\cloudtable
\edef\ltwosol{\pgfplotsretval}
\pgfplotstablegetelem{0}{Linf_sol}\of\cloudtable
\edef\linfsol{\pgfplotsretval}
  \begin{loglogaxis}[
    cloudplot,
    ylabel={Relative error},% compared to original $f^*$},
    legend pos=south east,
    legend style={row sep=0.7em, font=\large},
    cycle list name=exotic,
  ]
    \addplot+[can filtered] table[y expr=\thisrow{L1_error}/\lonesol] {cloud_wavelet_stats.csv} coordinate (lastrellone) {};
    \addlegendentry{plain}
    \addplot+[ds filtered] table[y expr=\thisrow{L1_error}/\lonesol] {cloud_wavelet_stats.csv};
    \addlegendentry{downsplit}
    \addplot+[can filtered] table[y expr=\thisrow{L2_error}/\ltwosol] {cloud_wavelet_stats.csv} coordinate (lastrelltwo) {};
    % \addlegendentry{$\frac{\norm{f - f^*}_2}{\norm{f^*}_2}$}
    \addplot+[can filtered] table[y expr=\thisrow{Linf_error}/\linfsol] {cloud_wavelet_stats.csv} coordinate (lastrellinf) {};
    % \addlegendentry{$\frac{\norm{f - f^*}_\infty}{\norm{f^*}_\infty}$}
    \addplot+[ds filtered] table[y expr=\thisrow{L2_error}/\ltwosol] {cloud_wavelet_stats.csv};
    \addplot+[ds filtered] table[y expr=\thisrow{Linf_error}/\linfsol] {cloud_wavelet_stats.csv};
  \end{loglogaxis}
  \node[normlabel, yshift=-4pt] at (lastrellone){$\frac{\norm{f - f^*}_1}{\norm{f^*}_1}$};
  \node[normlabel, yshift=15pt] at (lastrelltwo){$\frac{\norm{f - f^*}_2}{\norm{f^*}_2}$};
  \node[normlabel, yshift=25pt] at (lastrellinf){$\frac{\norm{f - f^*}_\infty}{\norm{f^*}_\infty}$};
\end{tikzpicture}
        \caption{Relative compression errors for $L_1, L_2$, and $L_\infty$ norms (double-log scale).
         Lossless compression introduces rounding errors less than \num{e-15}, outside the plotted range.}
        \label{fig:cloud:errors}
    \end{subfigure}\\[0.5em]
    \begin{subfigure}[t]{0.75\textwidth}
    \def\axisdefaultwidth{12cm}
\def\axisdefaultheight{7cm}

\pgfplotsset{
  constant/.style={
    domain=1e-10:2,
    mark=none,
    dashdotted
  },
  boxes/.style={
    table/y=boxes,
  },
}

\begin{tikzpicture}
  \pgfplotstableread[col sep=comma]{cloud/cloud_wavelet_stats.csv}\cloudtable

  \begin{semilogxaxis}[
    cloudplot,
    ymin=0,
    ylabel={Number of coefficients $N$},
    legend pos=south west,
    cycle list name=exotic,
    scaled y ticks=false,
    yticklabel style={/pgf/number format/fixed},
  ]
  \pgfplotstablegetelem{0}{vdb_orig_total_coefficients}\of\cloudtable
  \edef\vdborigcoeffs{\pgfplotsretval}
  \pgfplotstablegetelem{0}{boxes}\of\cloudtable
  \edef\losslesscancoeffs{\pgfplotsretval}
  \pgfplotstablegetelem{1}{boxes}\of\cloudtable
  \edef\losslessdscoeffs{\pgfplotsretval}
    \addplot[openvdb,constant] {\vdborigcoeffs};
    \addlegendentry{OpenVDB}
    \addplot+[plain,constant] {\losslesscancoeffs};
    \addlegendentry{plain lossless}
    \addplot+[downsplit,constant] {\losslessdscoeffs};
    \addlegendentry{downsplit lossless}
    \addplot+[can filtered,boxes] table {cloud_wavelet_stats.csv};
    \addlegendentry{plain lossy}
    \addplot+[ds filtered,boxes] table {cloud_wavelet_stats.csv};
    \addlegendentry{downsplit lossy}
  \end{semilogxaxis}
\end{tikzpicture}
        \caption{Number of remaining coefficients (semi-log scale)}
        \label{fig:cloud:coefficients}
    \end{subfigure}\\[0.5em]
    \begin{subfigure}[t]{0.75\textwidth}
        \provideboolean{showblosc}
        \setboolean{showblosc}{true}
        \provideboolean{showblosc}
\def\axisdefaultwidth{12cm}
\def\axisdefaultheight{7cm}

\pgfplotsset{
  invisible/.style={
    domain=1e2:2e2
  },
  constant/.style={
    domain=1e-10:2,
    mark=none,
    dashdotted
  },
  raw/.style={
    solid,
  },
  blosc/.style={
    densely dotted,
    thick,
    opacity=0.4,
  },
  desc/.style={
    mark=star,
  },
  values/.style={
    mark=square*,
  },
  total/.style={
    mark=*,
  },
  values raw/.style={
    raw,
    values,
    table/y expr=\thisrow{values_raw_bytes}/1048576,
  },
  values blosc/.style={
    blosc,
    values,
    table/y expr=\thisrow{values_blosc2_bytes}/1048576,
  },
  desc raw/.style={
    raw,
    desc,
    table/y expr=\thisrow{desc_raw_bytes}/1048576,
  },
  desc blosc/.style={
    blosc,
    desc,
    table/y expr=\thisrow{desc_blosc2_bytes}/1048576,
  },
  total raw/.style={
    raw,
    total,
    table/y expr=\thisrow{raw_total_bytes}/1048576,
  },
  total blosc/.style={
    blosc,
    total,
    table/y expr=\thisrow{blosc2_total_bytes}/1048576,
  },
}

\begin{tikzpicture}
  \pgfplotstableread[col sep=comma]{cloud/cloud_wavelet_stats.csv}\cloudtable

  \begin{semilogxaxis}[
    cloudplot,
    ymin=0,
    ymax=1.68,
    ylabel={File size},
    y unit={\si{\mebi\byte}},
    xlabel={Wavelet compression threshold $\varepsilon$},
    legend pos=north east,
    cycle list name=exotic,
    scaled y ticks=false,
    % y ticklabel style={/pgf/number format/fixed},
    legend pos=north west,
  ]
  \pgfplotstablegetelem{0}{vdb_orig_file_bytes}\of\cloudtable
  \edef\vdborigbytes{\pgfplotsretval}
  \pgfplotstablegetelem{0}{raw_total_bytes}\of\cloudtable
  \edef\losslesscanrawbytes{\pgfplotsretval}
  \pgfplotstablegetelem{0}{blosc2_total_bytes}\of\cloudtable
  \edef\losslesscanbloscbytes{\pgfplotsretval}
  \pgfplotstablegetelem{1}{raw_total_bytes}\of\cloudtable
  \edef\losslessdsrawbytes{\pgfplotsretval}
  \pgfplotstablegetelem{1}{blosc2_total_bytes}\of\cloudtable
  \edef\losslessdsbloscbytes{\pgfplotsretval}
    \addplot[openvdb,constant] {\vdborigbytes / 1048576};
    \addlegendentry{OpenVDB}

    %%%%% plots outside the visible range, to get the legend entries
    \addplot[plain,invisible,mark=none] {1};
    \addlegendentry{plain}
    \addplot[downsplit,invisible,mark=none] {1};
    \addlegendentry{downsplit}
    \addplot[constant,invisible] {1};\label{plot:lossless}
    \addplot[invisible,values] {1};\label{plot:values}
    \addplot[invisible,desc] {1};\label{plot:descriptor}
    \addplot[invisible,total] {1};\label{plot:total}
    \addplot[invisible,raw] {1};\label{plot:raw}
    \addplot[invisible,blosc] {1};\label{plot:blosc}
    %%%%%

    \addplot[plain,constant] {\losslesscanrawbytes / 1048576};
    \addplot[downsplit,constant] {\losslessdsrawbytes / 1048576};
    \addplot+[can filtered,values raw] table {cloud_wavelet_stats.csv};
    \addplot+[can filtered,desc raw] table {cloud_wavelet_stats.csv};
    \addplot+[can filtered,total raw] table {cloud_wavelet_stats.csv};
    \addplot+[ds filtered,values raw] table {cloud_wavelet_stats.csv};
    \addplot+[ds filtered,desc raw] table {cloud_wavelet_stats.csv};
    \addplot+[ds filtered,total raw] table {cloud_wavelet_stats.csv};

\ifthenelse{\boolean{showblosc}}{
    % \addplot+[plain,constant] {\losslessbloscbytes};
    \addplot+[can filtered,values blosc] table {cloud_wavelet_stats.csv};
    \addplot+[can filtered,desc blosc] table {cloud_wavelet_stats.csv};
    \addplot+[can filtered,total blosc] table {cloud_wavelet_stats.csv};
    \addplot+[ds filtered,values blosc] table {cloud_wavelet_stats.csv};
    \addplot+[ds filtered,desc blosc] table {cloud_wavelet_stats.csv};
    \addplot+[ds filtered,total blosc] table {cloud_wavelet_stats.csv};
}{}
    \coordinate(patternlegendpos) at (1e-5, 0.4);
    \coordinate(marklegendpos) at (0.15, 1.2);
  \end{semilogxaxis}
    \matrix[
    matrix of nodes,
    draw,
    inner sep=0.2em,
    % nodes={font=\footnotesize},
    fill=white,
  ]at(marklegendpos)
  {
    \ref*{plot:lossless} & lossless         & [5pt]\\
    \ref*{plot:total}  & total           & [5pt]\\
    \ref*{plot:values} & values         & [5pt]\\
    \ref*{plot:descriptor} & descriptor           & [5pt]\\
    };
% \ifthenelse{\boolean{showblosc}}{%-> weird catcode errors
  \matrix[
    matrix of nodes,
    draw,
    inner sep=0.2em,
    % nodes={font=\footnotesize},
    fill=white,
  ]at(patternlegendpos)
  {
    \ref*{plot:raw} & raw         & [5pt]\\
    \ref*{plot:blosc}  & \blosc{}           & [5pt]\\
    };
% }{}
\end{tikzpicture}
        \caption{File sizes (semi-log scale); total = values + descriptor.}
        \label{fig:cloud:storage}
    \end{subfigure}%TODO common horizontal axis?
    \caption{
       Lossy compression errors, coefficients, and storage for different wavelet thresholds, in comparison to the original OpenVDB data.
       The interesting range is $\varepsilon \in [\num{1e-2}, \num{3.3e-1}]$, where the lossy compression reduces the number of coefficients and file size significantly, while the visual quality is still good.
    %    Dashdotted constant lines show results for lossless compression.
    }
    \label{fig:cloud:errorcoeffstorage}
\end{figure*}
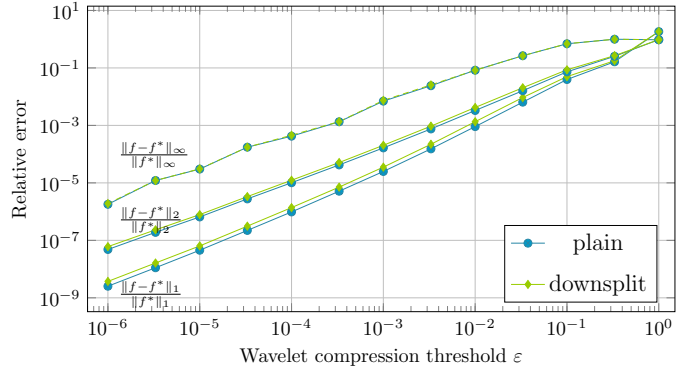
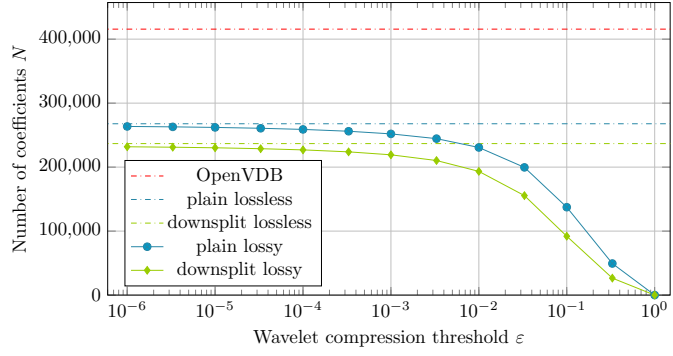
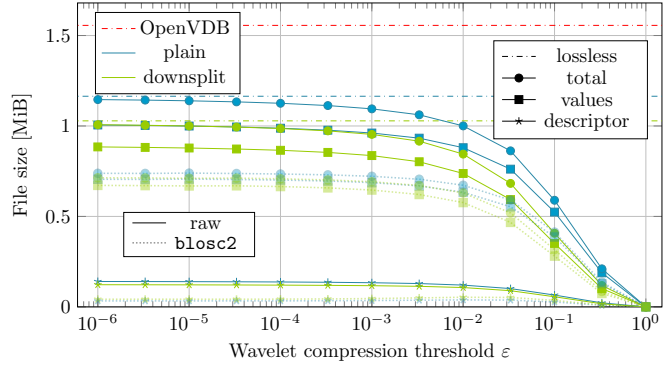

\Cref{fig:cloud:errors} plots the relative errors incurred by lossy compression with plain and downsplit coarsening.
The approximation error increases smoothly with the threshold $\varepsilon$.
As is to be expected, the downsplit-coarsening loop introduces slightly higher errors over plain coarsening for the same threshold, as more nodes can be merged under the same criterion.
The maximum threshold $\varepsilon = \num{1}$, fully coarsens the field to only the root node with the mean density of \num[round-precision=2]{0.04495194} everywhere, leading to a uniform cube in \cref{fig:cloud:vdb_blender}.
For all thresholds, the mass in the density field is conserved within \num{e-8}.
Lossless compression produces accurate results, with floating-point round-off staying below \num{e-15} (not shown).

In \cref{fig:cloud:coefficients}, the reduction in storage size is slow at first and becomes significant for $\varepsilon > \num{e-3}$.
Downsplit compression prunes away nearly a constant number of coefficients after plain coarsening, a difference that grows in relative importance as the storage size is further reduced.
At small $\varepsilon$, the difference in storage size between plain and downsplit coarsening is around $\qty[round-precision=0]{12.0483528}{\%}$.
At $\varepsilon = \num{3.3e-1}$, it is higher, around $\qty[round-precision=0]{46.2857143}{\%}$.
Notably, the $\varepsilon = \num{e-1}$ approximations, while retaining decent visual quality, need fewer than half the coefficients of their lossless counterparts and, in the case of downsplit, fewer than one quarter of the coefficients used by OpenVDB.

\Cref{fig:cloud:storage} plots the overall storage requirements in mebibytes, including both the data coefficients and the omnitree descriptor.
The overhead due to the tree descriptor is a small fraction of the total storage requirement, which is dominated by the real-valued data coefficients.
Thus, the total storage largely follows the number of coefficients as in \cref{fig:cloud:coefficients}.
One can also observe that lossless plain coarsening achieves a compression of \qty[round-precision=2]{1.336436029}{\xtimes} compared to OpenVDB, and downsplit raises this to \qty[round-precision=2]{1.513205608}{\xtimes}.
% Data is omitted for \blosc{} compression, since no clear recommendations can be drawn and the figure would have become too cluttered:
For the cloud data, we apply \blosc{} compression to both the data values and the descriptor.
The total added \blosc{} compression rates range from \num[round-precision=2]{1.285725} (for high $\varepsilon$ and downsplit) to \num[round-precision=2]{1.551752774} (for low $\varepsilon$ and plain coarsening). 
As a result, adding \blosc{} compression allows to achieve a storage footprint smaller than OpenVDB by \qty[round-precision=1]{2.148151659}{\xtimes} for lossless downsplit, and further reduce it to \qty[round-precision=1]{4.917732373}{\xtimes} with $\varepsilon = \num{e-1}$.
%, which means increasing the file size considerably  -> sorry, old values!
% These values would lead to largely overlapping plots with the raw omnitree storage in \cref{fig:cloud:storage}.
% \blosc{} compression has a larger effect on the descriptor than on the data coefficients, but since the latter dominate in the case of floating-point values, the influence of additional compression is less pronounced than for the \thingitenk{} data in \cref{fig:storage_boxplot}.
% \todo{stress more the added savings in the discussion and conclusion}

\section{Discussion}\label{sec:discussion}

The experiments detailed in \cref{sec:experiment} show that anisotropic structured compression can outperform the isotropic compression offered by OpenVDB in terms of both in-memory and on-disk storage.
How much exactly is saved depends both on the inherent (axis-aligned) anisotropy of the data, and the heuristic criteria employed for the coarsening and downsplit operations, which were the topic of \cref{sec:omnitree_operations}:
As long as only plain coarsening with Haar wavelets is employed, one needs to store significantly less data coefficients compared to OpenVDB, see \cref{fig:coeff_boxplot}, but the approximation rate still largely follows the quadtree complexity theorem.
However, if we combine the wavelet coarsening with tree transformations like downsplit, one observes an asymptotic increase of the approximation rate, meaning that the relative savings even increase with resolution.

In fact, the single-level-downsplit employed here is a simple heuristic that still misses many cross-level coarsening opportunities, so there is potential for further improvement through more refined tree transformation strategies.
% For the purposes of this work, we content ourselves with the experimental validation that downsplit is necessary to improve upon the theoretical expectation from the complexity theroem.
Analogously, coarseninng and downsplit by wavelet coefficients is just one among many possible heuristics, there are several ways in which this method could be refined, e.g., dynamic programming~\cite{thieleFastAlgorithmAdapted1996}.
(Conversely, the space-frequency tiling algorithm of Villemoes et al.~\cite{thieleFastAlgorithmAdapted1996,lindbergImageCompressionAdaptive2000} may benefit from omnitree data structures, particularly in higher dimensionalities.)

Furthermore, adding \blosc{} compression was evaluated for on-disk storage of omnitree descriptors, showing some compression benefits for larger omnitrees.
Additional compression with \blosc{} appears more effective for plain coarsening at \qty[round-precision=1]{2.527695962}{\xtimes}, but still sees some savings for downsplit coarsening at \qty[round-precision=1]{1.39038014}{\xtimes}.
For continuous-valued data, the potential for omnitree compression is tightly linked to the errors one wants to admit in the scheme, and \blosc{} adds consistent savings of \qtyrange{1.29}{1.55}{\xtimes} on top.
It might be interesting to evaluate other approaches to scientific data compression, such as SZ~\cite{liangSZ3ModularFramework2023}, in particular for the downsplit data.
Similarly, the omnitree multiscale framework could be interesting for further research into mixed-precision computation and storage schemes:
For fine-scale details, lower precision is usually sufficient, while the coarser features operate in higher precision, analogously to RAPTOR's mixed-precision experiments~\cite{hoeroldRAPTORPracticalNumerical2025}.

Compared to OpenVDB objects, however, it is fair to say that lossless compression by large factors (up to \qty[round-precision=1]{35.81620162}{\xtimes} for coefficients and \qty[round-precision=1]{27.6145849495733}{\xtimes} for file sizes) was possible in our experiments.
A runtime comparison is contingent on a high-performance omnitree implementation, and out of the scope of this work.
In fact, the authors believe that the hierarchy implemented in OpenVDB could be a good fit to make omnitrees a high-performance data structure, for example by a forest-of-omnitrees-of-omnitrees-of-omnitrees... approach.
The tree hierarchy could have a fixed depth, and segment the tree by different wavelet thresholds, allowing for adaptive reconstruction despite irregular access patterns.
This could also map cleanly to distributed-memory approaches as implemented in \texttt{p4est}~\cite{bursteddeP4estScalableAlgorithms2011} and \texttt{t8code}~\cite{holkeT8codeModularAdaptive2025}.

Our method conserves the first moment (mass) during compression by using Haar wavelets.
Higher conservation orders are possible with higher orders of wavelets, for example CDF lifting wavelets~\cite{cohenBiorthogonalBasesCompactly1992} and Alpert mutliwavelets~\cite{alpertClassBasesSparse1993}.
The discontinuous Alpert multiwavelets in particular could be a great fit for omnitrees, as their supports are nonoverlapping per level (and Haar wavelets are in fact their lowest-order instantiation).
Combining them with the ultraweak DG approaches in $hp\mathrm{3D}$~\cite{chakrabortyAnisotropicHpadaptationFramework2024a,hennekinghp3DScalableMPI2024} could yield very effective multiscale schemes.
But even for the smoother CDF wavelets with larger supports, AMM~\cite{bhatiaAMMAdaptiveMultilinear2022} illustrates how \enquote{stamping} can be used to transform the hierarchical wavelet representation exactly to a nodal representation for continuous biorthogonal wavelets.
After stamping, the result can then be used for visualization and other workflow steps that operate on non-hierarchical (nodal) values.
This could be valuable for applications that require the numerical stability usually hard to obtain with interpolating adaptive schemes:
Many interpolating schemes are similar or equivalent to interpolet / lazy wavelet basis compression, which is inherently numerically unstable~\cite{kosterMultiskalenbasierteFiniteDifferenzenVerfahrenAuf2002,pollingerStableMassconservingSparse2023}.

Equipped with the algorithms to refine and coarsen omnitrees, it is now possible to develop fully dynamic AMR schemes that could be particularly beneficial to higher-dimensional simulations, i.e., 4-6 spatial dimensions.
These dimensionalities are common for high-fidelity simulations of plasma microturbulence with the Vlasov-Boltzmann equation, where current approaches often rely on structured and block-structured grids.
Omnitrees or block-omnitree data structures have the potential to allow for fully adaptive spatial refinement in this area of scientific computing for the first time.

\section{Conclusion}\label{sec:conclusion}
This work shows that omnitrees can be effectively coarsened and compressed, and significant advantages are enabled by the omnitree downsplit operation.
Wavelet representations fit well into the omnitree framework, as they allow to coarsen by thresholding the detail coefficients, and also give a heuristic for guiding the downsplit operation.
This was validated on binary-valued data from the \thingitenk{} data set, and continuous-valued data from the Disney cloud data set, showing significant savings in terms of number of coefficients and storage size compared to OpenVDB.
In summary, omnitrees can outperform the expectations of the quadtree/octree complexity theorem in storage, and further improvements are possible through more refined heuristics for coarsening and downsplit, as well as through efficient iteration and parallelization of the data structure.

\newpage

\section*{Acknowledgments}
We thank Stefan Henneking, Nancy Hitschfeld Kahler, Tobias Weinzierl, and Stefan Zellmann for discussing their previous work on similar data structures with us.
We thank Prof. Gerald Pao for pointing us to the work of Villemoes et al.~\cite{thieleFastAlgorithmAdapted1996,lindbergImageCompressionAdaptive2000} regarding omnitree-like tiling in the space/frequency domain.
We thank Stefan Zimmer for suggesting the scale-independent coarsening criterion (\cref{eq:coarsening_criterion}).

\section*{Declaration of Generative AI and AI-assisted technologies in the writing process}
Statement: During the preparation of this work the authors used Claude and Microsoft Copilot in order to edit the manuscript text, and Claude Code to assist with code generation.
After using this tool/service, the authors reviewed and edited the content as needed and take full responsibility for the content of the publication.

\section*{Reproducibility References}
Code used for this work includes omnitree data structures~\cite{bleifreiFreifrauvonbleifreiDyAda2025} and function approximation as well as tree transformations using wavelets~\cite{bleifreiFreifrauvonbleifreiWavelets_with_omnitrees2026}; both are available through github under GPL-3.0.
Intermediate data produced are available through a research data repository\footnote{will be published after review; preview link: \url{https://zenodo.org/records/19659695?preview=1&token=eyJhbGciOiJIUzUxMiJ9.eyJpZCI6Ijc1NzAwNGQ0LWQwOWQtNDMzOS05NDlkLTE3YWFhYzA0NDQ4NiIsImRhdGEiOnt9LCJyYW5kb20iOiI2ZmY2ZjcwOTYwY2RmNDdiZWQ5MjMwZGMxODAyZjlhMyJ9.xp3B0567kIMBkgU41xmE_hx-5Hfv3dw9xAhwnFQJqCcjPYGCjY9J50xT79qIqRm5n31q0loO3CRmvJEMUHKLtA}}. %~\cite{}.%TODO

\ifthenelse{\boolean{arxiv}}{
    \printbibliography
}{
    \bibliography{bib}
}

\end{document}